\documentclass[pra,aps,floatfix,twocolumn,superscriptaddress,amsmath,amssymb,]{revtex4-2}
\usepackage{graphicx}
\usepackage{color}
\usepackage{bm}
\usepackage{hyperref}
\usepackage[normalem]{ulem}
\hypersetup{colorlinks=true,citecolor=blue,linkcolor=blue,urlcolor=blue}
\allowdisplaybreaks

\begin{document}

\title{R\'{e}nyi entanglement entropy after a quantum quench
starting from insulating states in a free boson system}

\author{Daichi Kagamihara}
\email{kagamihara@phys.kindai.ac.jp}
\thanks{These authors contributed equally to this work.}
\affiliation{%
Department of Physics, Kindai University, Higashi-Osaka, Osaka 577-8502, Japan}
\affiliation{%
Department of Physics, Chuo University, Bunkyo, Tokyo 112-8551, Japan}

\author{Ryui Kaneko}
\email{rkaneko@phys.kindai.ac.jp}
\thanks{These authors contributed equally to this work.}
\affiliation{%
Department of Physics, Kindai University, Higashi-Osaka, Osaka 577-8502, Japan}

\author{Shion Yamashika}
\affiliation{%
Department of Physics, Chuo University, Bunkyo, Tokyo 112-8551, Japan}

\author{Kota Sugiyama}
\affiliation{%
Department of Physics, Chuo University, Bunkyo, Tokyo 112-8551, Japan}

\author{Ryosuke Yoshii}
\affiliation{%
Center for Liberal Arts and Sciences, Sanyo-Onoda City University, Yamaguchi 756-0884, Japan}

\author{Shunji Tsuchiya}
\affiliation{%
Department of Physics, Chuo University, Bunkyo, Tokyo 112-8551, Japan}

\author{Ippei Danshita}
\affiliation{%
Department of Physics, Kindai University, Higashi-Osaka, Osaka 577-8502, Japan}

\date{\today}

\begin{abstract}
We investigate the time-dependent R\'{e}nyi entanglement entropy
after a quantum quench starting from the Mott-insulating
and charge-density-wave states in a one-dimensional free boson system.
The second R\'{e}nyi entanglement entropy is found to be the negative of the logarithm
of the permanent of a matrix consisting
of time-dependent single-particle correlation functions.
From this relation and a permanent inequality, we obtain
rigorous conditions for satisfying the volume-law entanglement growth.
We also succeed in calculating the time evolution of the R\'{e}nyi entanglement entropy
in extremely large systems by
brute-force computations of the permanent. 
We discuss possible applications of our findings
to the real-time dynamics of noninteracting bosonic systems.
\end{abstract}

\maketitle

\section{Introduction}

The concept of entanglement is indispensable for
understanding quantum many-body physics these days.
A pure quantum many-body state is entangled
when it cannot be represented by a product state~\cite{nielsen2010}.
The entanglement entropy quantifies the degree of entanglement
and is a valuable probe for characterizing states of quantum many-body systems.
For example,
in critical systems,
the entanglement entropy exhibits the universal scaling
with the size of a subsystem;
the universal coefficient is determined by
the corresponding conformal field
theory~\cite{holzhey1994,calabrese2004,calabrese2009,vidal2003,eisert2010,laflorencie2016}.
Topologically ordered states,
which cannot be described by conventional order parameters,
would be characterized by the topological entanglement
entropy~\cite{kitaev2006,levin2006,zhang2011,isakov2011}.

The von Neumann entanglement entropy is a standard reference value
to quantify the entanglement.
When a system possesses a pure state $|\psi\rangle$
and can be divided into two subsystems A and B,
the von Neumann entanglement entropy is defined as
$S_{\mathrm{vN}}=-\mathrm{Tr}_{\mathrm{A}} \hat{\rho}_{\mathrm{A}} \ln \hat{\rho}_{\mathrm{A}}$,
where $\hat{\rho}_{\mathrm{A}} = \mathrm{Tr}_{\mathrm{B}} \hat{\rho}$ is the reduced density matrix
of $\hat{\rho}=|\psi\rangle\langle\psi|$ and $\mathrm{Tr}_{\mathrm{A(B)}}$ is the trace over the basis of subsystem A (B).
The R\'{e}nyi entanglement entropy is another quantity,
which behaves similarly to the von Neumann entanglement
entropy~\cite{zyczkowski2003,daley2012},
and is defined as
$S_{\alpha}=[\ln\mathrm{Tr}_{\mathrm{A}} (\hat{\rho}_{\mathrm{A}}^{\alpha})]/(1-\alpha)$.
The von Neumann entanglement entropy can be regarded as
the $\alpha \rightarrow 1$ limit of the R\'{e}nyi entanglement entropy.

The entanglement entropy is not merely an ideal quantity in theory,
but it is also measurable experimentally.
The protocol for measuring the R\'{e}nyi entanglement entropy
was first proposed in 2012~\cite{abanin2012,daley2012}.
In Ref.~[\onlinecite{daley2012}],
the authors considered the real-time dynamics of cold atoms
in optical lattices,
which can be realized in experiments~\cite{greiner2002}, 
and suggested preparing two copies of the same state.
The R\'{e}nyi entanglement entropy can be evaluated by controlling
the tunnel coupling between these copies
and by measuring the parity of the atom numbers.
The second R\'{e}nyi entanglement entropy has been indeed observed 
during a quench dynamics of one-dimensional (1D) cold atomic gases
in an optical lattice~\cite{islam2015,kaufman2016}.

Previous theoretical studies have actively discussed the dynamics of entanglement entropy after a quantum quench in connection with information propagation and thermalization~\cite{calabrese2005,eisert2010,laflorencie2016,
yoshii2022,dechiara2006,bardarson2012,bauer2015,goto2019,lauchli2008,kunimi2021,
rylands2022,
alba2017,yao2020,
fagotti2008,
yamashika2022_arxiv}.
While the dynamics of entanglement entropy in 1D lattice systems can be accurately analyzed by means of numerical methods based on matrix product states~\cite{dechiara2006,bardarson2012,bauer2015,goto2019,lauchli2008,kunimi2021}, the tractable time scale is rather limited in general due to the linear growth of the entanglement entropy in time~\cite{calabrese2005}.
In the case of 1D systems described by fermionic quasiparticles, such as the transverse-field Ising model, when the system is quenched to a noninteracting parameter region,
long-time dynamics can be investigated
analytically~\cite{alba2018,calabrese2005,fagotti2008,frerot2015}.
In particular, when the initial state is a Gaussian state,
i.e., a ground or thermal state of some free (quadratic) Hamiltonian,
the time-evolved state remains Gaussian.
Then, the time evolution of the entanglement entropy
can be evaluated from single-particle correlation functions.
The fermionic Gaussian states include simple product states
such as the Mott-insulating (MI) and charge-density-wave (CDW) states,
which can also be prepared in experiments.

By contrast, studies of the entanglement growth during the quench dynamics for (soft-core)
bosons are very few even in the case that the system is quenched to the noninteracting point.
This is partly because a simple product state, which is often used as an initial state of quench dynamics in  experiments~\cite{islam2015,kaufman2016}, is not a Gaussian state for bosonic systems.
Starting from product states such as the MI and CDW states, single-particle correlation functions are analytically calculable~\cite{cramer2008a,barmettler2012,cheneau2012}.
However, it is not straightforward to calculate the entanglement entropy from these correlation functions because a time-evolved state is not a Gaussian state.

This situation raises the following questions:
(i) Can we get an analytical form of the R\'{e}nyi entanglement entropy 
concerning real-time dynamics
of noninteracting bosonic systems?
(ii) Supposing we obtain the analytical form,
can we rigorously obtain conditions under which the volume-law scaling
is satisfied during the real-time evolution?
(iii) Can we numerically evaluate the R\'{e}nyi entanglement entropy in systems
much larger than the best currently available methods
can handle?

To answer these questions,
we take the 1D soft-core
Bose-Hubbard model as the simplest playground.
When the system is quenched to the noninteracting Hamiltonian,
we obtain the analytical form of evaluating the second R\'{e}nyi
entanglement entropy.
It can be expressed by the expectation value of the shift (\textsc{swap})
operator~\cite{daley2012,abanin2012,islam2015,kaufman2016}
and is given as the negative of the logarithm
of the permanent of a time-dependent matrix
consisting of single-particle correlation functions.
We also give the condition for the volume-law scaling of the R\'{e}nyi
entanglement entropy
by using a permanent inequality~\cite{berkowitz2018}.
In addition, we obtain the long-time evolution of the R\'{e}nyi
entanglement entropy
in extremely large systems by numerically computing the matrix permanent.
Although direct calculations of the permanent require
exponential-time cost in general,
accessible sizes are found to be much larger than
the exact diagonalization and matrix-product-state methods can deal with.
Last but not least, we propose that
the infinity norm of rows of the matrix consisting of the correlation function
offers an entropy-density-like value
and would give a practical bound for the R\'{e}nyi entanglement entropy.
This value would free us from exponential-time computations
of the permanent, as long as we are interested in a qualitative behavior
of the entanglement entropy growth rather than the value itself.

This paper is organized as follows:
In Sec.~\ref{sec:1d_bose_hubbard},
we present the 1D Bose-Hubbard model
and introduce two initial states for the quench dynamics.
In Sec.~\ref{sec:eval_renyi},
we calculate the time-evolved states after the sudden quench
and derive the analytical form of the R\'{e}nyi entanglement entropy.
In Sec.~\ref{sec:analytical_results},
we summarize some interesting properties of the matrix consisting of the correlation function
introduced in Sec.~\ref{sec:eval_renyi}
and show the condition for the volume-law scaling of the R\'{e}nyi
entanglement entropy.
We describe some examples of the application of the condition to quench
dynamics in our model.
In Sec.~\ref{sec:numerical_results},
we directly compute the permanent of the matrix consisting of the correlation function
and obtain the time evolution of the R\'{e}nyi entanglement entropy.
We also compare our results with other reference values.
In addition, we introduce an entropy-density-like value, which can be calculated 
in polynomial time,
and discuss a bound for the R\'{e}nyi entanglement entropy.
In Sec.~\ref{sec:summary},
we draw our conclusions
and discuss possible applications to several problems on
real-time dynamics of free boson systems.
Throughout this paper, we set $\hbar = 1$,
take the lattice constant to be unity,
and consider the zero-temperature dynamics, for simplicity.

\section{One-dimensional Bose-Hubbard Model}
\label{sec:1d_bose_hubbard}

We consider the quench dynamics in the 1D Bose-Hubbard model
under the open boundary condition.
The Hamiltonian is defined as
\begin{align}
 \hat{H} &= - J \sum_{j=1}^{L-1} ( \hat{b}^{\dagger}_j \hat{b}_{j+1} + \mathrm{H.c.} )
 + \sum_{j=1}^{L} \Omega_j \hat{n}_j
\nonumber
\\
 &~\phantom{=}~
 + \frac{U}{2} \sum_{j=1}^{L} \hat{n}_j (\hat{n}_j-1).
\end{align}
Here the symbols $\hat{b}_j$ and $\hat{n}_j$ denote
the boson annihilation and number operators, respectively.
The strength of the hopping and the interaction
are given as $J$ and $U$, respectively, and $\Omega_j$ denotes the single-particle potential.
The number of sites is represented as $L$. This model quantitatively describes 1D Bose gases in optical lattices when the lattice potential is sufficiently deep.

We focus on the quench from insulating states
to the noninteracting ($U=0$) and homogeneous ($\Omega_j = 0$) point.
As initial states,
we specifically choose the MI state at unit filling,
which is represented as
\begin{align}
 |\psi^\mathrm{MI}\rangle = \prod_{j=1}^{L} \hat{b}^{\dagger}_j |0\rangle,
\end{align}
and the $010101\cdots$ CDW state at half filling,
which is described as
\begin{align}
 |\psi^\mathrm{CDW}\rangle = \prod_{j = 2,4,\dots}^{L} \hat{b}^{\dagger}_{j} |0\rangle,
\end{align}
where $|0\rangle$ is the vacuum state of $\hat{b}_j$ and $L$ is taken as an even number.
The MI state is the ground state of the Bose-Hubbard model at unit
filling for the large-$U$ limit and can be prepared in experiments via a
slow ramp-up of the optical lattice
potential~\cite{cheneau2012,islam2015,kaufman2016,takasu2020}.
The CDW state is the ground state of the Bose-Hubbard model at half
filling when $\Omega_j = \Omega(-1)^{j+1}$, $\Omega/J \gg 1$, and $U / J \gg 1$. It can
be prepared in experiments with use of a secondary optical lattice whose
lattice constant is twice as large as that of the primary
lattice~\cite{trotzky2012}.

\section{Evaluating the second R\'{e}nyi entanglement entropy using shift operators}
\label{sec:eval_renyi}

We first consider the time evolution of the many-body wave function.
Since the matrix representation of the single-particle Hamiltonian after the quench is tridiagonal, we easily find the single-particle energy
\begin{align}
 \epsilon_k
 &=
 - 2J \cos \left( \frac{k\pi}{L+1} \right)
\end{align}
and corresponding eigenstate
\begin{align}
x_{k,l}
 &=
 \sqrt{\frac{2}{L+1}} \sin \left( \frac{k\pi}{L+1} l \right),
 \label{eq:X_definition}
\end{align}
where $k,l = 1, 2,\dots,L$.
The time-evolved many-body states
$|\psi(t)\rangle = e^{ -i \hat{H} t} |\psi(0)\rangle$
are given as
\begin{align}
 |\psi^{\mathrm{MI}}(t)\rangle
 &=
 \prod_{j=1}^{L}
 \left[ \sum_{l=1}^{L} y_{j,l}(t) \hat{b}^{\dagger}_l \right]
 |0\rangle,
\\
 |\psi^{\mathrm{CDW}}(t)\rangle
 &=
 \prod_{j=2,4,\dots}^{L}
 \left[ \sum_{l=1}^{L} y_{j,l}(t) \hat{b}^{\dagger}_l \right]
 |0\rangle,
\end{align}
where all information about real-time dynamics is encoded in
\begin{align}
 y_{j,l}(t)
 =
 \sum_{k=1}^{L} x_{k,j} e^{-i\epsilon_k t} x_{k,l}.
 \label{eq:Y_definition}
\end{align}

The second R\'{e}nyi entanglement entropy can be obtained by utilizing the expectation value
of the shift (\textsc{swap}) operator
$\hat{V}$~\cite{daley2012,abanin2012,islam2015,kaufman2016}.
Let us suppose that we have two copies of the state $|\psi(t)\rangle$,
which we call copies 1 and 2, and that the total wave function is given by the product state of the two copies,
\begin{align}
|\psi_{\mathrm{copy}}(t)\rangle = |\psi(t) \rangle \otimes |\psi(t) \rangle.
\label{eq:total_wave_func}
\end{align}
We divide the system into two subsystems A and B.
Here subsystem A contains $j=1,2,\dots, L_{\mathrm{A}}$ sites in this paper.
Let us consider the shift operator $\hat{V}_{\mathrm{A}}$ which swaps states in subsystem A.
The expectation value of $\hat{V}_{\mathrm{A}}$ in terms of $|\psi_{\mathrm{copy}}(t)\rangle$ is related to the reduced density matrix $\hat{\rho}_{\mathrm{A}}$ as
\begin{align}
\langle \psi_{\mathrm{copy}}(t) | \hat{V}_{\mathrm{A}} |\psi_{\mathrm{copy}}(t)\rangle
 = \mathrm{Tr}'_{\mathrm{A}} (\hat{\rho}_{\mathrm{A}} \otimes \hat{\rho}_{\mathrm{A}} \hat{V}_{\mathrm{A}})
 = \mathrm{Tr}_{\mathrm{A}} \hat{\rho}_{\mathrm{A}}^2,
\label{eq:relation_shift_op_density_matrix}
\end{align}
where $\mathrm{Tr}'_{\mathrm{A}}$ stands for the trace over the basis of
subsystem A of copies 1 and 2.
As long as $\hat{V}_{\mathrm{A}}$ acts on a product state of copies 1 and 2, such as Eq.~(\ref{eq:total_wave_func}),
the shift operator transforms the creation operator as
\begin{align}
\hat{V}_{\mathrm{A}} \hat{b}^{\dag}_{j} \hat{V}_{\mathrm{A}}^{-1}
&=
\begin{cases}
\hat{c}^{\dag}_{j} & (j \in \mathrm{A})
\\
\hat{b}^{\dag}_{j} & (j \in \mathrm{B}),
\end{cases}
\label{eq:action_shift_operator_1}
\\
\hat{V}_{\mathrm{A}} \hat{c}^{\dag}_{j} \hat{V}_{\mathrm{A}}^{-1}
&=
\begin{cases}
\hat{b}^{\dag}_{j} & (j \in \mathrm{A})
\\
\hat{c}^{\dag}_{j} & (j \in \mathrm{B}),
\end{cases}
\label{eq:action_shift_operator_2}
\end{align}
where operators $\hat{b}$ and $\hat{c}$ respectively correspond to boson operators for copies $1$ and $2$.
For derivation of Eqs.~(\ref{eq:action_shift_operator_1}) and
(\ref{eq:action_shift_operator_2}), see Appendix~\ref{sec:action_shift_operator}.
Making use of these relations, we can evaluate the second R\'enyi
entanglement entropy in an elementary way.
For example, for the MI state,
it can be evaluated as
\begin{align}
 S_2
 &=
 - \ln \langle \psi^{\mathrm{MI}}_{\mathrm{copy}}(t)|
 \hat{V}_{\mathrm{A}} | \psi^{\mathrm{MI}}_{\mathrm{copy}}(t)\rangle,
\\
 |\psi^{\mathrm{MI}}_{\mathrm{copy}}(t)\rangle
 &=
 \left\{
 \prod_{j=1}^{L}
 \left[ \sum_{l=1}^{L} y_{j,l}(t) \hat{b}^{\dagger}_l \right]
 \right\}
\nonumber
\\
 &~\phantom{=}~\times
 \left\{
 \prod_{j=1}^{L}
 \left[ \sum_{l=1}^{L} y_{j,l}(t) \hat{c}^{\dagger}_l \right]
 \right\}
 |0\rangle^{\otimes 2},
\\
 \hat{V}_{\mathrm{A}} |\psi^{\mathrm{MI}}_{\mathrm{copy}}(t)\rangle
 &=
 \left\{
 \prod_{j=1}^{L}
 \left[
 \sum_{l=1}^{L_{\mathrm{A}}} y_{j,l}(t) \hat{c}^{\dagger}_l
 +
 \sum_{l=L_{\mathrm{A}}+1}^{L} y_{j,l}(t) \hat{b}^{\dagger}_l
 \right]
 \right\}
\nonumber
\\
 &\!\!\!\!\!\!\!\!\!\!\!\!\times
 \left\{
 \prod_{j=1}^{L}
 \left[
 \sum_{l=1}^{L_{\mathrm{A}}} y_{j,l}(t) \hat{b}^{\dagger}_l
 +
 \sum_{l=L_{\mathrm{A}}+1}^{L} y_{j,l}(t) \hat{c}^{\dagger}_l
 \right]
 \right\}
 |0\rangle^{\otimes 2}.
\end{align}

Since both $|\psi^{\mathrm{MI}}_{\mathrm{copy}}(t)\rangle$ and $\hat{V}_{\mathrm{A}} |\psi^{\mathrm{MI}}_{\mathrm{copy}}(t)\rangle$ are many-boson states and their wave functions are symmetric under the permutation of $\hat{b}$ bosons (and $\hat{c}$ bosons as well),
the expectation value
$\langle \psi_{\mathrm{copy}}(t)| \hat{V}_{\mathrm{A}} | \psi_{\mathrm{copy}}(t)\rangle$
can be rewritten by the permanent of 
single-particle correlation functions $Z$ and $\tilde{Z}$:
\begin{align}
\langle \psi_{\mathrm{copy}}(t)| \hat{V}_{\mathrm{A}} | \psi_{\mathrm{copy}}(t)\rangle = \mathrm{perm} \begin{pmatrix} \tilde{Z} & Z \\ Z & \tilde{Z} \end{pmatrix},
\end{align}
where $Z$ $(\tilde{Z})$ is a single-particle overlap matrix between
single-particle states 
$e^{-i\hat{H}t} \hat{b}^{\dag}_j |0\rangle^{\otimes 2}$ and 
$\hat{V}_{\mathrm{A}} e^{-i\hat{H} t} \hat{c}^{\dag}_{l} |0 \rangle^{\otimes 2}$
($e^{-i\hat{H}t} \hat{b}^{\dag}_j |0\rangle^{\otimes 2}$ and 
$\hat{V}_{\mathrm{A}} e^{-i\hat{H} t} \hat{b}^{\dag}_{l} |0 \rangle^{\otimes 2}$).
Using the fact that $Z + \tilde{Z} = I$, where $I$ is an identity matrix, we obtain the analytical expression of the R\'{e}nyi entanglement entropy:
\begin{align}
 S_2 &= -\ln \mathrm{perm} A_Z,
 \label{eq:entropy_permanent_formula}
\\
 A_Z &=
 \begin{pmatrix}
 I-Z & Z \\
 Z & I-Z \\
 \end{pmatrix}.
 \label{eq:A_definition}
\end{align}
Here the element $z_{j,l}$ of the Hermitian matrix $Z$ is given as
\begin{align}
\label{eq:def_zij}
 z^{\mathrm{MI}}_{j,l}
 &=
 \sum_{m=1}^{L_{\mathrm{A}}} y^{*}_{j,m}(t) y_{l,m}(t)
 \quad
 (j,l=1, 2, \dots, L),
\\
 z^{\mathrm{CDW}}_{j,l}
 &=
 \sum_{m=1}^{L_{\mathrm{A}}} y^{*}_{2j,m}(t) y_{2l,m}(t)
 \quad
 (j,l=1, 2, \dots, L/2).
\end{align}
Note that a somewhat similar formula for the R\'{e}nyi entanglement entropy given by the
permanent has been proposed for excited states in the static
system~\cite{zhang2021}.

In the following, $\tilde{L}$ denotes the size of the square matrix $A_Z$.
It is given by $2N$, where $N$ is the number of particles.
For example, $\tilde{L} = $ $2L$ ($L$) for the MI (CDW) state.
Hereafter we will mainly consider the R\'{e}nyi entanglement entropy
for a bipartition of the system into two half chains ($L_{\mathrm{A}}=L/2$).

\section{Analytical results}
\label{sec:analytical_results}

In this section,
we analytically evaluate the system size $L$ dependence of
the R\'{e}nyi entanglement entropy $S_2$
by examining the permanent of the matrix $A_Z$.
We will discuss the condition for $S_2$
to satisfy the volume-law scaling.
Then, we apply the obtained volume-law condition to the quench
dynamics of the present case.

\subsection{Remarks on the matrices \texorpdfstring{$Z$}{Z} and \texorpdfstring{$A_Z$}{AZ}}
\label{sec:remarks_on_matrix_Z}

Let us first summarize the characteristics of the matrices $Z$ and $A_Z$.
The purpose of this section is to show $||A_Z||_2 = 1$ for both MI and CDW initial states.
Here
the operator $2$-norm is defined as
$||A||_2 := \sup_{||\bm{x}||_2\le 1, \bm{x} \in \mathbb{C}^{M}} ||A\bm{x}||_2$ with $M$ being the size of a square matrix $A$
[on the right hand side of the equation,
$||\bm{x}||_p := (\sum_j |x_j|^p)^{1/p}$ is an $L^p$ norm
of a vector $\bm{x}$],
or the largest singular value of the matrix $A$.
We will utilize this fact to obtain the condition for the volume-law
entanglement growth in the next section.

For the quench from the MI state, the matrix $Z^{\mathrm{MI}}$ is
a complex orthogonal projection matrix,
satisfying $(Z^{\mathrm{MI}})^2 = Z^{\mathrm{MI}} = (Z^{\mathrm{MI}})^{\dagger}$
(see Appendix~\ref{sec:more_matrix_z}).
Therefore, all the eigenvalues are either $0$ or $1$.
For the quench from the CDW state,
the matrix $Z^{\mathrm{CDW}}$ is
a principal submatrix of the Hermitian matrix $Z^{\mathrm{MI}}$; i.e.,
it can be obtained from $Z^{\mathrm{MI}}$
by removing $L/2$ rows and the same $L/2$ columns.
Using Cauchy's interlace theorem~\cite{hwang2004,fisk2005},
we can show that all the eigenvalues of $Z^{\mathrm{CDW}}$
are bounded by the largest eigenvalue $1$
and the smallest eigenvalue $0$ of $Z^{\mathrm{MI}}$.
As a result, $0 \le ||Z^{\mathrm{CDW}}||_2 \le ||Z^{\mathrm{MI}}||_2 = 1$.

All the eigenvalues of the matrix $A_Z$ can be obtained from those of $Z$.
Let us write the eigenvalues of $Z$ as $\epsilon^{(Z)}_k$
and the eigenvectors of $Z$ as
$|\epsilon^{(Z)}_k\rangle = (v_{k,1},v_{k,2},\dots,v_{k,n})^T$
for $k=1,2,\dots,n$
with $n(=\tilde{L}/2)$ being the length of the square matrix $Z$.
Then, half of all the eigenvalues of $A_Z$ are $\epsilon^{(A_Z)}_k = 1$,
and the corresponding eigenvectors are
$|\epsilon^{(A_Z)}_k\rangle
= (v_{k,1},v_{k,2},\dots,v_{k,n},v_{k,1},v_{k,2},\dots,v_{k,n})^T$.
The remaining half are $\epsilon^{(A_Z)}_{k+n} = 2\epsilon^{(Z)}_k - 1$,
and the corresponding eigenvectors are
$|\epsilon^{(A_Z)}_{k+n}\rangle
= (v_{k,1},v_{k,2},\dots,v_{k,n},-v_{k,1},-v_{k,2},\dots,-v_{k,n})^T$.
Therefore, $\epsilon^{(A_Z)}_k, \epsilon^{(A_Z)}_{k+n}\in [-1,1]$ because $\epsilon^{(Z)}_k\in [0,1]$.
Thus, the operator $2$-norm of the matrix $A_Z$ satisfies $||A_Z||_2 = 1$.

For the quench from the MI state,
the matrix $A_Z^{\mathrm{MI}}$ becomes a unitary matrix,
which can be shown by the relations
$(Z^{\mathrm{MI}})^2 + (I-Z^{\mathrm{MI}})^2
= Z^{\mathrm{MI}} + (I-Z^{\mathrm{MI}}) = I$
and
$Z^{\mathrm{MI}} (I-Z^{\mathrm{MI}}) = 0$.
This unitarity also ensures $||A^{\mathrm{MI}}_Z||_2 = 1$.
For the quench from the CDW state, the matrix $A^{\mathrm{CDW}}_Z$ is not a unitary matrix
in general; however, $||A^{\mathrm{CDW}}_Z||_2 = 1$ still holds.

Note that the elements of matrix $A_Z$ satisfy
\begin{align}
 \sum_{l} a_{j,l} = 1, \quad
 \sum_{j} a_{j,l} = 1,
\end{align}
for any rows $j$ and columns $l$,
which is a part of the definition of the doubly stochastic matrix
while the nonnegativity condition $a_{j,l} \geq 0$ is absent.
(Here the matrix $A_Z$ is complex and satisfies $0\le |a_{j,l}| \le \max(|a_{j,l}|) \le ||A_Z||_2 = 1$.)
The permanent of the doubly stochastic matrix has been intensively
studied~\cite{marcus1960,wilf1966,merris1973,friedland1979,minc1984,cheon2005,laurent2010,gurvits2014},
while little is known about the permanent of a general complex matrix so far.

\subsection{Condition for volume-law entanglement entropy}
\label{subsec:volume_law}

To quantify the $\tilde{L}$ dependence of the entanglement entropy,
we utilize the inequality~\cite{berkowitz2018}
\begin{align}
\label{eq:permA_ineq}
 |\mathrm{perm} A|
 \le
C^{M} \exp \left\{- 10^{-5} \times \left[ 1-\frac{g_{A}(M)}{C} \right]^2 M \right\},
\end{align}
which holds for an arbitrary $M\times M$ complex matrix $A$
and an arbitrary nonzero constant $C$ satisfying
$C\ge ||A||_2$.
The function $g_{A}(M)$
is defined as
\begin{align}
g_{A}(M) := \frac{1}{M} \sum_{j=1}^{M} ||\bm{r}_j||_{\infty}
\end{align}
with $\bm{r}_j$'s being rows of a matrix $A$
and $||\bm{x}||_{\infty} := \max_j |x_j|$, which satisfies $0 \leq g_{A}(M) \leq ||A||_2$.
\par
We apply inequality~(\ref{eq:permA_ineq}) to our case given by Eqs.~(\ref{eq:entropy_permanent_formula}) and (\ref{eq:A_definition}).
From Eq.~(\ref{eq:relation_shift_op_density_matrix}), $\mathrm{perm} A_Z$ equals $\mathrm{Tr}_{\mathrm{A}} \hat{\rho}_{\mathrm{A}}^2$, implying $\mathrm{perm} A_Z > 0$.
In addition,
because $||A_Z||_2=1$, we can choose $C=1$
as the tightest bound.
Then, the inequality is simplified as
\begin{align}
 \mathrm{perm} A_Z
 \le
 \exp \left\{- 10^{-5} \times \left[ 1-g_{A_Z}(\tilde{L}) \right]^2 \tilde{L} \right\},
\end{align}
where
\begin{align}
g_{A_Z}(\tilde{L})
= \frac{1}{N} \sum_{j=1}^{N} \max_{l=1}^{N} ( |z_{j,l}|,|\delta_{j,l} - z_{j,l}| ).
\label{eq:g_A_Z_definition}
\end{align}
Note that $|\mathrm{perm} A|\le 1$ holds for any unitary
matrix $A$~\cite{marcus1962},
as is the case with the quench from the MI state.
Even if $A$ is nonunitary, as is the case with the quench from the CDW state,
$|\mathrm{perm} A|\le (|| A ||_2)^{M}$ holds~\cite{Gurvits2005}.
Inequality~(\ref{eq:permA_ineq}) gives a much tighter constraint on the permanent of $A$ than these two inequalities.

Consequently, the R\'{e}nyi entanglement entropy satisfies
\begin{align}
\label{eq:s2_ineq}
S_2 \ge 10^{-5} \times \left[ 1-g_{A_Z}(\tilde{L}) \right]^2 \tilde{L}.
\end{align}
This inequality rigorously guarantees that when
\begin{align}
\label{eq:condition_volume_law}
\lim_{\tilde{L} \rightarrow\infty} \left[ 1 - g_{A_Z}(\tilde{L}) \right] \neq 0,
\end{align}
the R\'{e}nyi entanglement entropy shows the volume-law scaling.
In other words, when Eq.~(\ref{eq:condition_volume_law}) holds, the area-law scaling (and the area-law scaling with a logarithmic correction, as is often the case in critical systems~\cite{holzhey1994,calabrese2004,calabrese2009,vidal2003}) is prohibited.
If $\lim_{L\rightarrow\infty} [1 - g_{A_Z}(\tilde{L}) ] = 0$
($\tilde{L} \propto L$), inequality~(\ref{eq:s2_ineq}) becomes meaningless and either the area-law or volume-law scaling is allowed.
From the volume-law condition given in Eq.~(\ref{eq:condition_volume_law}), we expect that
the value $1-g_{A_Z}(\tilde{L})~(\geq 0)$
could be used as an entropy-density-like value, which we will discuss in Sec.~\ref{subsec:relation_inf_norm}.

\subsection{Application to the quench dynamics}
Here we give two examples which violate the volume-law condition in Eq.~(\ref{eq:condition_volume_law}).
One is a product state and the other is a time-evolved state after a short time.
Neither of these
is expected
to follow the volume-law scaling, and in the following, we show that they indeed break the
condition in Eq.~(\ref{eq:condition_volume_law}).
Here, we take a state starting from the MI state as an example.

The first case is a product state
at $t=0$, which apparently has zero entanglement entropy.
This fact implies the violation of the condition in Eq.~(\ref{eq:condition_volume_law}).
A straightforward calculation on the matrix $Z^{\mathrm{MI}}$
gives a matrix
\begin{align}
Z^{\mathrm{MI}} = \begin{pmatrix}
I_{N/2} & 0_{N/2}
\\
0_{N/2} & 0_{N/2}
\end{pmatrix}.
\label{eq:z_MI_t_equals_zero}
\end{align}
The matrix $A_Z$ 
becomes just a 
permutation matrix, which is a square matrix whose every row and column contains
a single $1$ with $0$s elsewhere.
In fact, $g_{A_Z}(\tilde{L}) = ||A_Z||_2 = 1$ 
if and only if
the matrix $A_Z$ is a permutation matrix~\cite{berkowitz2018}.
Therefore, the product state always breaks the condition irrespective of $L$.
On the other hand, we can directly calculate the R\'{e}nyi entanglement entropy by the permanent.
The permanent of a permutation matrix is 
unity, 
and thus
the entanglement entropy for a product state is zero, as expected.

The second case is a time-evolved state with a short time, $t \ll L / v_{\mathrm{C}}$, where $v_{\mathrm{C}}$ is the propagation velocity of correlations.
In the present 1D free boson system, $v_{\mathrm{C}}$ is 
equivalent to
the maximum quasiparticle velocity $v_{\mathrm{max}}$.
The velocity is given by the maximal group velocity
and
$v_{\mathrm{max}}=2J$~\cite{cheneau2012,barmettler2012} (when $\hbar=1$ and the lattice constant is chosen to be unity)
in the present case.
Because the single-particle correlation function extends up to a distance $v_{\mathrm{C}} t$,
it is likely that
the entanglement entropy
does not
grow with increasing $L$ when $L > v_{\mathrm{C}} t$.
Therefore, the time-evolved state with fixed $t (\ll L / v_{\mathrm{C}})
$ is expected to follow the area-law scaling, implying the breaking of
the volume-law condition in Eq.~(\ref{eq:condition_volume_law}).

\begin{figure}[!t]
\centering
\includegraphics[width=\columnwidth]{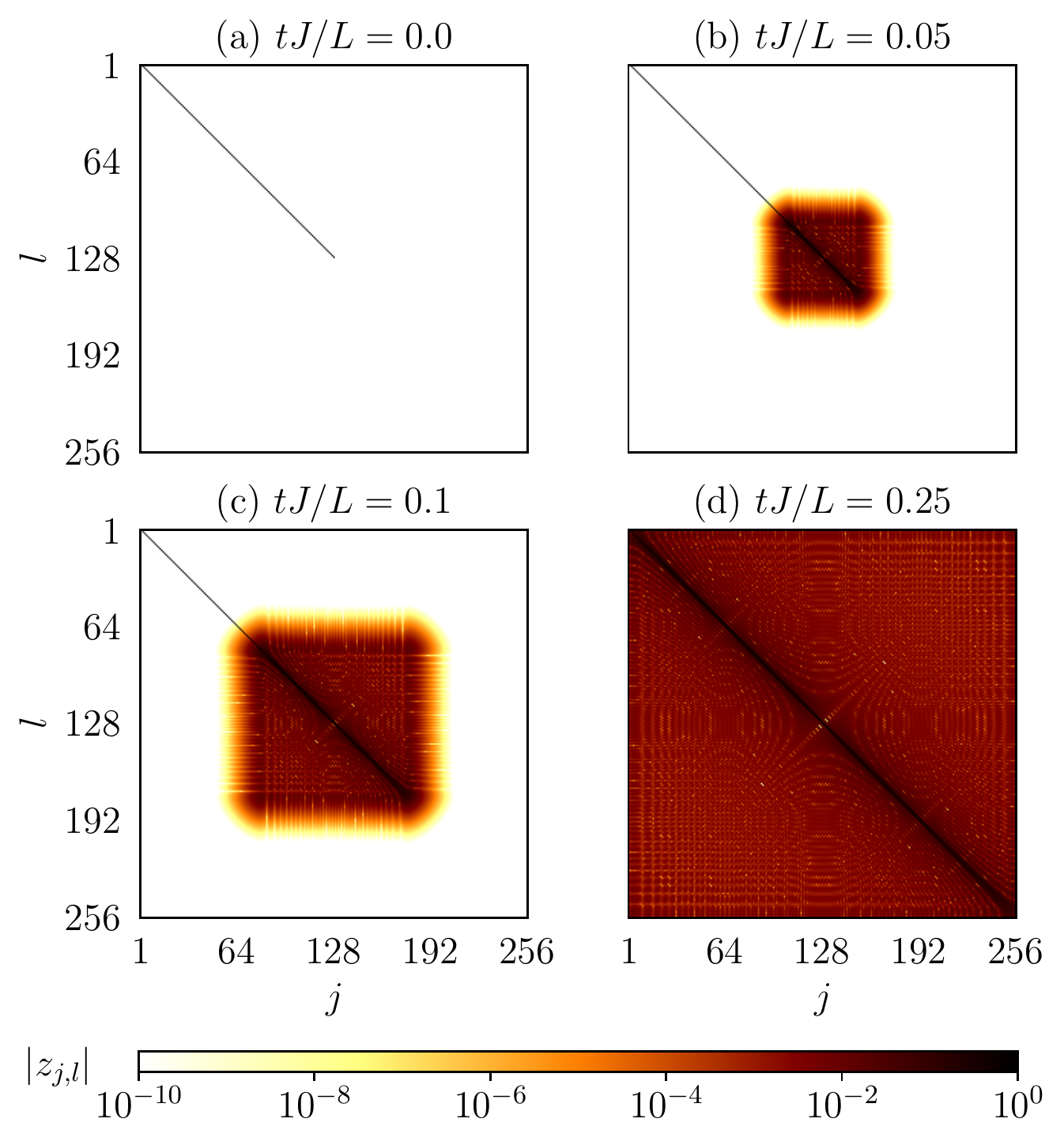}
\caption{
Time dependence of the absolute value of the correlation function $|z_{j,l}|$ from the MI initial state with $L = 256$
at (a) $tJ/L=0$, (b) $tJ/L=0.05$, (c) $tJ/L=0.1$, and (d) $tJ/L=0.25$.
At $t = 0$, the correlation function $z_{j,l}$ is given by Eq.~(\ref{eq:z_MI_t_equals_zero}).
For $0 < tJ < L/4$, $z_{j,l}$ is approximately given by Eq.~(\ref{eq:z_approximation}) and the length of the sides of the square where $|z_{j,l}| > 10^{-10}$ holds is nearly equal to $2\beta \approx 4tJ$.
}
\label{fig:abs_z_time_dependence}
\end{figure}

In this situation, we can approximate the matrix $Z^{\mathrm{MI}}$ as
\begin{align}
Z^{\mathrm{MI}} \approx \begin{pmatrix}
I_{N/2 - \beta} & 0 & 0 \\
0 & Z'_{2\beta} & 0 \\
0 & 0 & 0_{N/2 - \beta}
\end{pmatrix},
\label{eq:z_approximation}
\end{align}
where $\beta$ is a positive integer and
is roughly proportional to $2tJ$.
$Z'_{2\beta}$ would be a dense matrix with the size $2\beta \times 2\beta$.
Figure~\ref{fig:abs_z_time_dependence} shows the time dependence of the absolute value of $Z^{\mathrm{MI}}$.
We see that $Z^{\mathrm{MI}}$ actually has the matrix structure given in Eq.~(\ref{eq:z_approximation}) when $t J \ll L/4$.
For the derivation of Eq.~(\ref{eq:z_approximation}) and the specific value of $\beta$, see Appendix~\ref{sec:breaking_volume_law}.

Using Eq.~(\ref{eq:z_approximation}), we can verify the breaking of the volume-law condition of Eq.~(\ref{eq:condition_volume_law}).
Substituting Eq.~(\ref{eq:z_approximation}) into $g_{A_Z}(\tilde{L})$ in Eq.~(\ref{eq:g_A_Z_definition}), we obtain
\begin{align}
g_{A_Z}(\tilde{L}) =& \frac{N - 2 \beta}{N} + \frac{1}{N} \sum_{j \in \Delta}
\max_{l \in \Delta} ( |z_{j,l}|,|\delta_{j,l} - z_{j,l}| ).
\end{align}
where $\Delta = [N/2 - \beta, N/2 + \beta]$.
Because $0 \leq |z_{j,l}| \leq 1$, $1- g_{A_Z}(\tilde{L})$ is bounded like
\begin{align}
0 \leq 1- g_{A_Z}(\tilde{L}) \leq \frac{2\beta}{N}.
\end{align}
Since $\beta$ depends on $2tJ$ but not on the system size $L$, by taking the thermodynamic limit $L(\propto N) \to \infty$, we conclude 
\begin{align}
\lim_{L \to \infty} \left[ 1 - g_{A_Z}(\tilde{L}) \right] = 0.
\end{align}

We note that the breaking of the volume-law condition does not directly mean the area-law scaling, as mentioned before.
Because we do not have the criterion on the area-law scaling of the entanglement entropy at this stage, 
we rely on the numerical calculation
to check the area-law scaling.
The numerical results of the R\'{e}nyi entanglement entropy will be shown in the next section.
Calculated R\'{e}nyi entanglement entropies with a short time $tJ \ll L$ 
shown in Fig.~\ref{fig:s2_bf} take almost the same value with increasing $L$, implying that the entanglement entropy of this state would obey the area-law scaling.

Another less rigorous evidence of the area-law entanglement scaling can be seen from the permanent formula.
Substituting the approximated expression on $Z^{\mathrm{MI}}$ in Eq.~(\ref{eq:z_approximation}) into Eq.~(\ref{eq:A_definition}), we obtain the permanent of the matrix $A_Z$ as
\begin{align} \nonumber
\mathrm{perm} A_Z &\approx \mathrm{perm} A_{Z'}
\\
&=
\mathrm{perm} \begin{pmatrix}
I_{2\beta} - Z'_{2\beta} & Z'_{2\beta}
\\
Z'_{2\beta} & I_{2\beta} - Z'_{2\beta}
\end{pmatrix}.
\end{align}
Applying the same discussion in Sec.~\ref{subsec:volume_law} with replacing the matrix size $\tilde{L}$ with $2\beta$, 
we expect that the entanglement entropy would be constant when $1-g_{A_{Z'}} = 0$ or proportional to $2\beta$ when $1-g_{A_{Z'}} \neq 0$.
Strictly speaking, this argument does not exclude the possibility that the entanglement entropy is also proportional to the system size $L$.
However, since the condition for the volume-law scaling in Eq.~(\ref{eq:condition_volume_law}) itself ensures the proportionality of the
entanglement entropy with respect to $\tilde{L}$, it is unlikely that the volume-law scaling of entanglement would be satisfied if the proportionality to $\tilde{L}$ is replaced by that to $\beta$.

\section{Numerical results}
\label{sec:numerical_results}

In this section, we will numerically evaluate the permanent to obtain
the R\'{e}nyi entanglement entropy.
In general, permanent calculations require an exponentially long time.
However, we can practically obtain the R\'{e}nyi entanglement entropy for systems
larger than the exact diagonalization method can handle
and can perform longer simulations than the method based on matrix product states,
even by performing a brute-force permanent calculation.

The advantages of getting the R\'{e}nyi entanglement entropy by the permanent calculation
are the following:
(i) We do not need Hamiltonian eigenstates, which require much memory cost.
This is the main reason why our method enables us to access larger systems than the exact diagonalization method can handle.
(ii) Without explicit time evolution,
we can directly calculate the R\'{e}nyi entanglement entropy at a given time,
which allows us parallel computations.
(iii)
In a system of soft-core bosons,
there is no upper limit to the number of bosons at any site,
but it is common to set a realistic limit when performing
numerical calculations.
With the present method we have proposed,
we do not have to care about the size of such a local Hilbert space limitation.

\subsection{Summary of algorithm}

Here we briefly review the algorithm for the permanent calculation.
The permanent of an $M\times M$ matrix $A$ is defined as
\begin{align}
 \mathrm{perm} A
 =
 \sum_{\sigma\in {\rm Sym}(M)}
 \prod_{j=1}^{M} a_{j,\sigma(j)},
\end{align}
where ${\rm Sym}(M)$ is the symmetric group,
i.e., over all permutations of numbers $1$, $2$, $\dots$, $M$.
Since straightforward calculations require $M! \times M$ arithmetic operations, we should use a more efficient algorithm.
The best known algorithms so far are the Ryser formula~\cite{ryser1963,brualdi1991,glynn2010} and Balasubramanian-Bax-Franklin-Glynn (BBFG) formula~\cite{balasubramanian1980,bax1996,bax1998,glynn2010,glynn2013}.
Both take $\mathcal{O}(M 2^{M-1})$ computation time.
Hereafter we mainly use the BBFG formula for the permanent calculation.
It is given by
\begin{align}
 \mathrm{perm} A
 =
 \frac{1}{2^{M-1}}
 \sum_{\bm{\delta}}
 \left(\prod_{m=1}^{M} \delta_m\right)
 \prod_{l=1}^{M} \sum_{j=1}^{M} \delta_j a_{j,l},
\end{align}
where $\bm{\delta}=(\delta_1,\delta_2,\dots,\delta_{M}) \in \{\pm 1\}^M$
with $\delta_1=1$.
Although the computation time using the above straightforward BBFG formula is $\mathcal{O}(M^2 2^{M-1})$, 
it can be reduced to $\mathcal{O}(M 2^{M-1})$ by utilizing a specific ordering of the binary numeral system, known as Gray code~\cite{gray1953,nijenhuis1978}.
Our numerical source code is based on the \textsc{python} program in
``The Walrus'' library~\cite{walrus}.

The current feasible matrix size is up to
$\sim 50\times 50$~\cite{neville2017,wu2018,lundow2022}.
For the quench from the MI (CDW) state, the size of the matrix $A_Z$
is $2L\times 2L$ ($L\times L$).
Therefore, in principle, we may handle $L\lesssim 25$ ($L\lesssim 50$)
for the MI (CDW) case.
Here we present our results for $L\le 20$ ($L\le 40$)
for the quench from the MI (CDW) state.
Although we may be able to utilize the symmetry of the matrix $A_Z$
to accelerate permanent computations~\cite{chin2018},
we stick to the original BBFG formula.
Even without improving the original algorithm,
it allows us to calculate the permanent for systems much larger than the exact diagonalization method can deal with.
(Note that, for example, exact diagonalization calculations for
$L=14$ from the MI state to $U/J=3.01$ have been reported~\cite{goto2019}.)
As for the size at which the permanent is computable,
the R\'{e}nyi entanglement entropy can be obtained at any given time,
allowing for longer simulations than
the methods based on matrix product states.

\subsection{Time dependence of R\'{e}nyi entanglement entropy}

\begin{figure}[!t]
\centering
\includegraphics[width=\columnwidth]{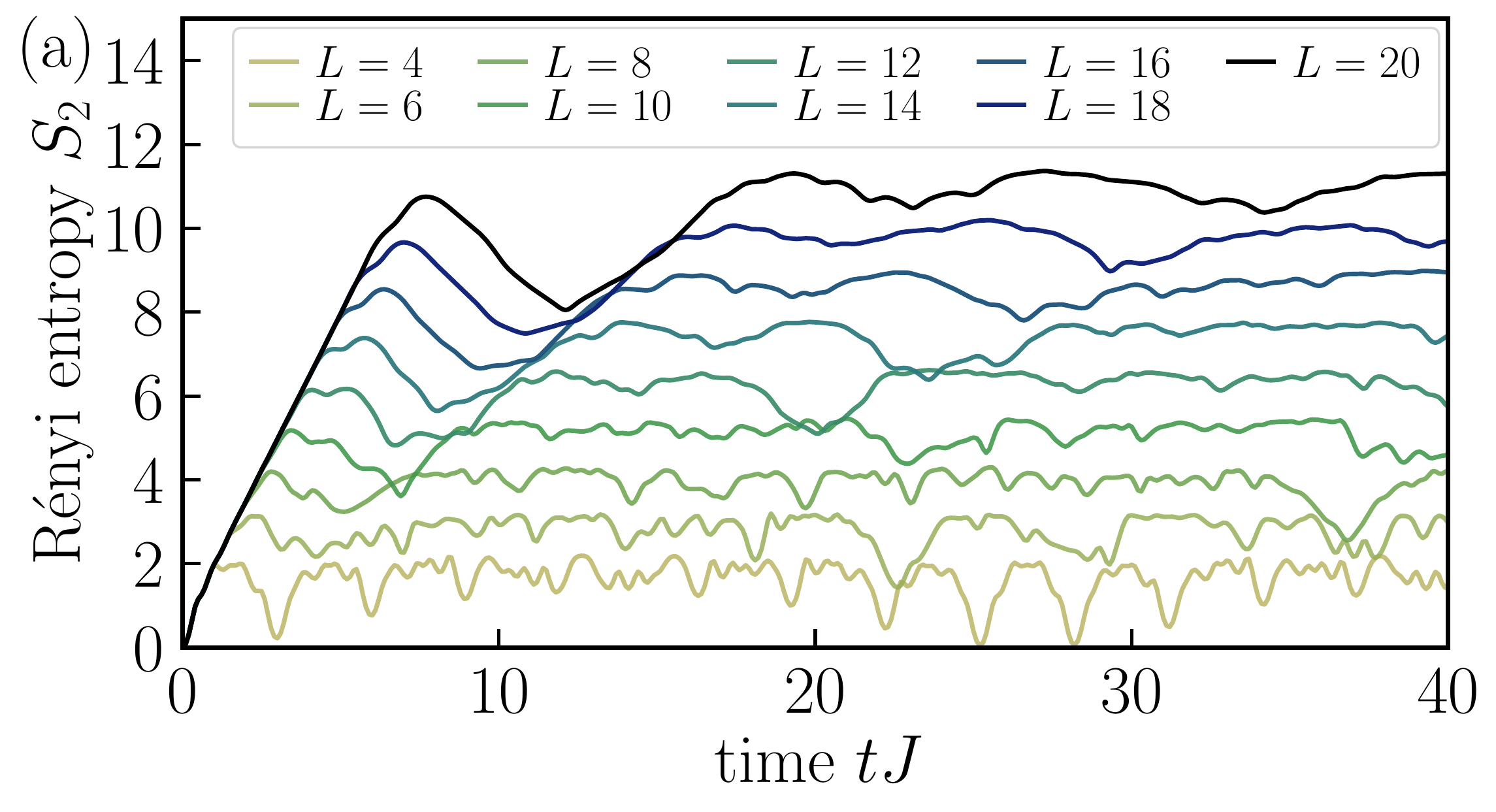}
\\
\includegraphics[width=\columnwidth]{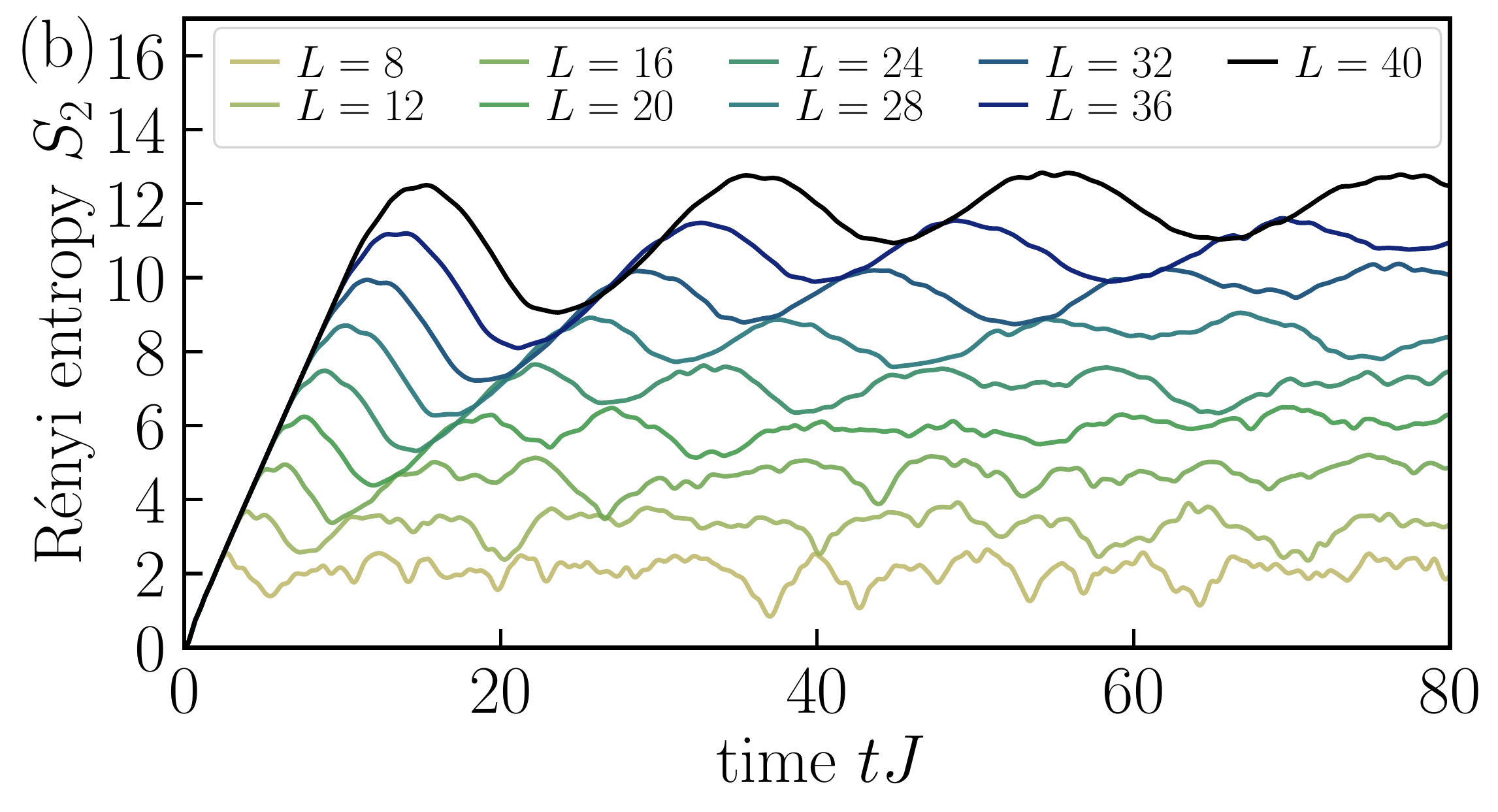}
\caption{Time dependence of the R\'{e}nyi entanglement entropy
for the quench from the (a) MI and (b) CDW states.
For a short time ($tJ\lesssim L/4$),
the R\'{e}nyi entanglement entropy $S_2$ exhibits an increase proportional to time $t$.
After a long time ($tJ\gtrsim L/4$),
$S_2$ is nearly saturated at the value
proportional to the system size $L$.}
\label{fig:s2_bf}
\end{figure}

\begin{figure}[!t]
\centering
\includegraphics[width=\columnwidth]{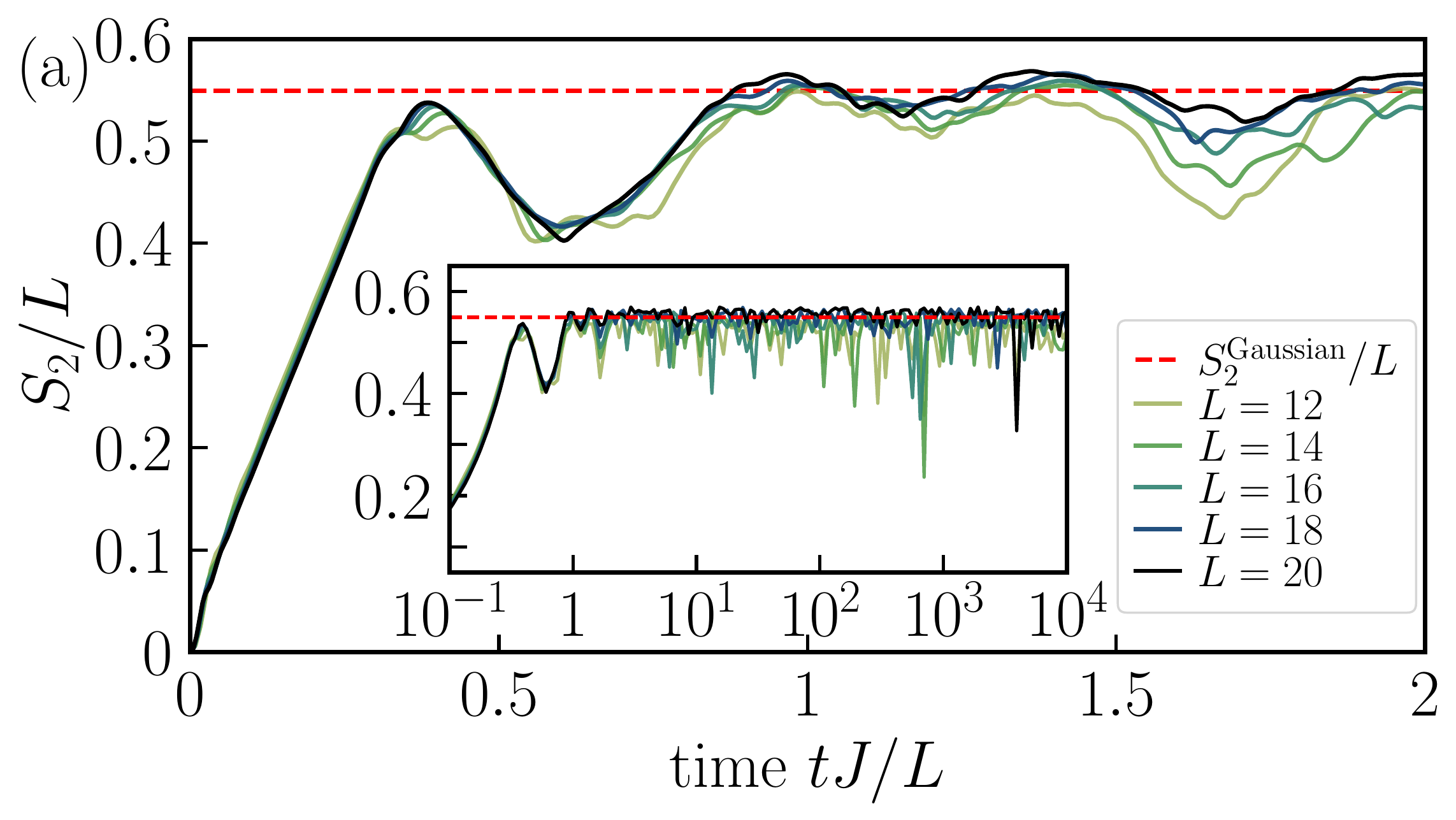}
\\
\includegraphics[width=\columnwidth]{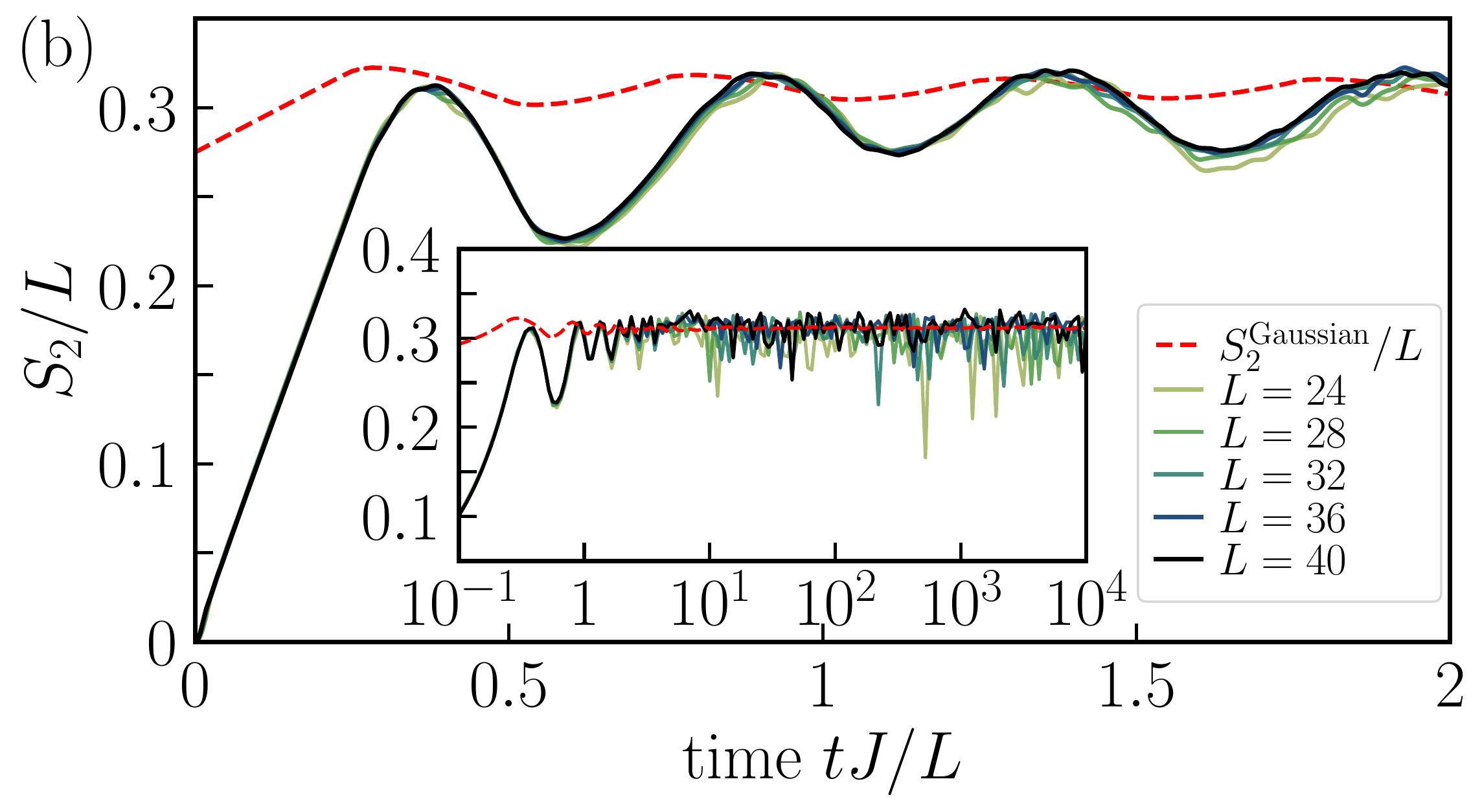}
\caption{Rescaled time dependence of the R\'{e}nyi entanglement entropy
for the quench from the (a) MI and (b) CDW states.
The R\'{e}nyi entanglement entropy and the time are rescaled by $L$.
The dashed lines correspond to the R\'{e}nyi entanglement entropy densities $S_2^{\rm Gaussian}/L$ estimated
from Eq.~(\ref{eq:gaussian_Renyi}).
The dashed line is independent of $L$ for the MI state,
while it is obtained for a sufficiently large size ($L=1024$)
for the CDW state.
The insets show the long-time behavior of the R\'{e}nyi entanglement entropy densities.
}
\label{fig:s2_div_L_bf}
\end{figure}

We examine the time dependence of the R\'{e}nyi entanglement entropy
(see Fig.~\ref{fig:s2_bf}).
For a short time ($tJ\lesssim L/4$), $S_2$ grows linearly with $t$.
After $tJ\gtrsim L/4$,
$S_2$ is almost saturated
and reaches the value nearly proportional to $L$.
$S_2$ exhibits oscillations whose period grows with $L$.
These observations are consistent with the fact that
the $t$-linear growth of the entanglement entropy terminates at $t\sim L/(2v_{\mathrm{max}})$,
where $v_{\mathrm{max}} = 2J$ is the maximum quasiparticle velocity~\cite{alba2017,cheneau2012,barmettler2012}.

To see this behavior more clearly,
we rescale the time $tJ$ and the R\'{e}nyi entanglement entropy $S_2$
in the unit of the system size $L$
(see Fig.~\ref{fig:s2_div_L_bf}).
All the lines nearly overlap for $tJ/L \lesssim 1$
when $L\gtrsim 10$ ($L\gtrsim 20$)
for the quench from the MI (CDW) state.
The deviation from the thermodynamic limit appears to be smaller
for the CDW state because the feasible size is larger than the MI state.

\subsection{Comparison with the R\'{e}nyi entanglement entropy
estimated from the Gaussian state}

\begin{figure}[!t]
\centering
\includegraphics[width=\columnwidth]{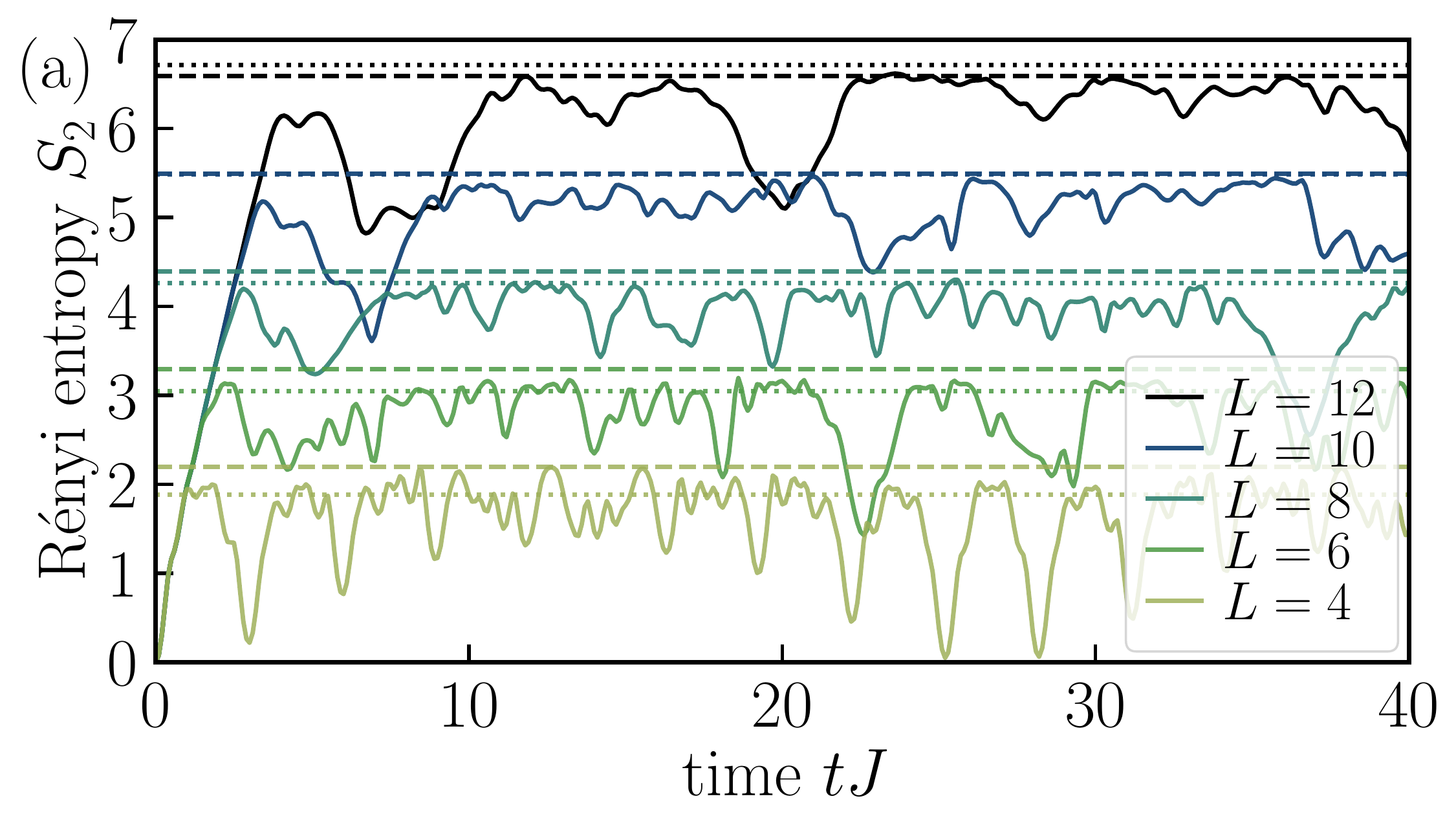}
\\
\includegraphics[width=\columnwidth]{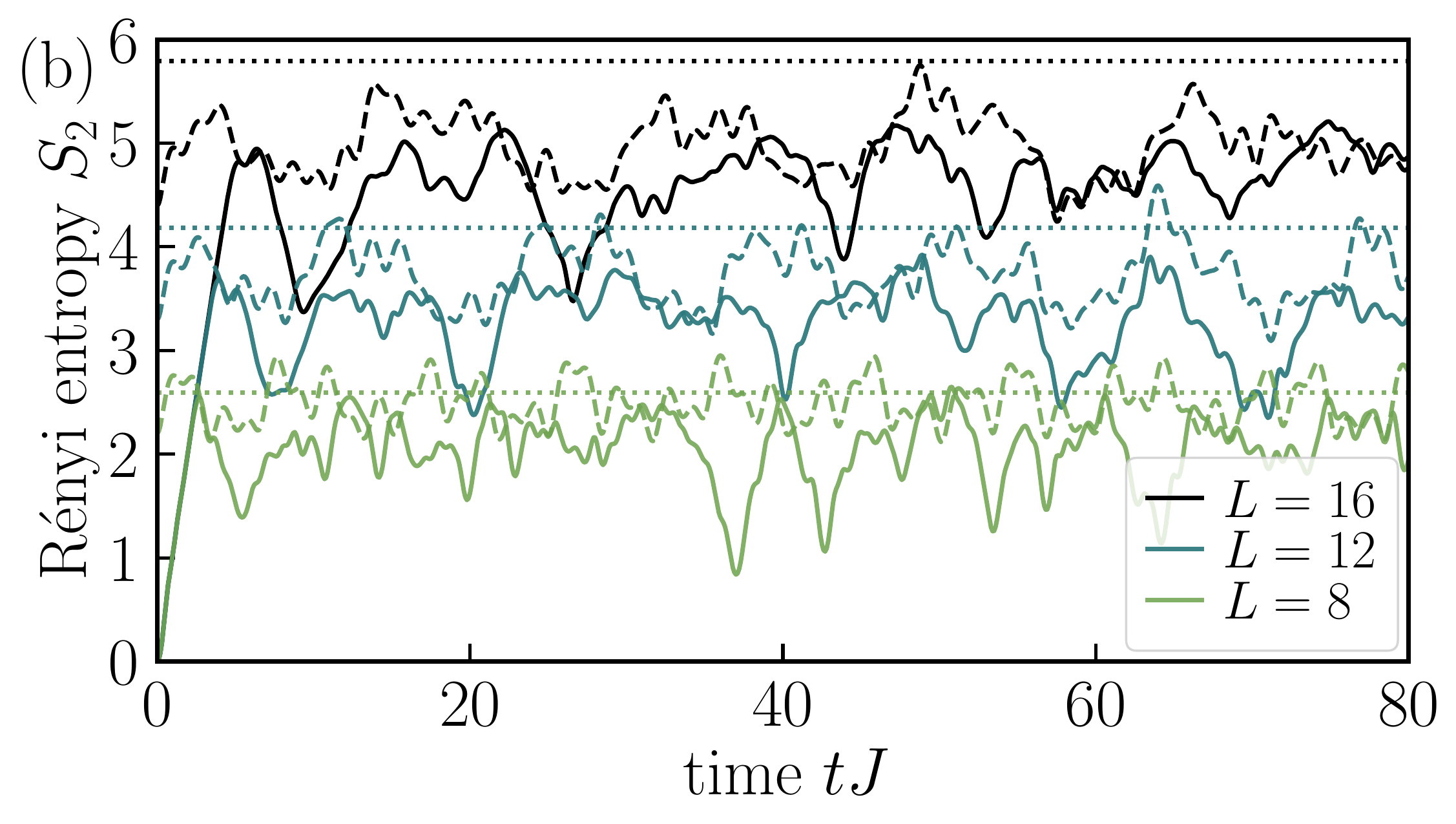}
\caption{%
Comparisons among the R\'{e}nyi entanglement entropy $S_2$ (solid lines),
that estimated from Eq.~(\ref{eq:gaussian_Renyi}), $S_2^{\rm Gaussian}$ (dashed lines), and the Page value $S_2^{\mathrm{Page}}$ (dotted lines)
for the (a) MI and (b) CDW states.
Note that $S_2^{\mathrm{Gaussian}}$ and $S_2^{\mathrm{Page}}$ of $L=10$ for the MI state take almost the same value.
We do not show error bars of $S_2^{\mathrm{Page}}$ because they are invisibly small.
}
\label{fig:s2_cmp_gaussian}
\end{figure}

The MI or CDW state quenched
to $U=0$ evolves to a Gaussian state
after a long time in the thermodynamic limit
($L \gg tJ \gg 1$)~\cite{cramer2008a,flesch2008,cramer2008b}.
The entanglement entropy of the Gaussian state
can be calculated from the eigenvalues of the matrix,
consisting of $\langle \hat{b}^{\dagger}_j \hat{b}_l \rangle$ and
$\langle \hat{b}_j \hat{b}_l \rangle$~\cite{alba2018,calabrese2005,fagotti2008,frerot2015,frerot2016}.
After diagonalizing the matrix in bosonic systems,
we obtain the eigenvalues, which correspond to
the expectation values of the mode occupation numbers $n_{\mu}$.
The R\'{e}nyi entanglement entropy $S_{\alpha}$ of order $\alpha$ can be described
by $n_{\mu}$ as~\cite{alba2018,calabrese2005,fagotti2008,frerot2015,frerot2016}
\begin{align}
 S_{\alpha}
 &=
 \frac{1}{\alpha-1}
 \sum_{\mu} \ln \left[(n_{\mu}+1)^{\alpha} - n_{\mu}^{\alpha}\right].
 \label{eq:gaussian_Renyi}
\end{align}

The time-evolved state is not a Gaussian state for $tJ, L<\infty$ in general.
However, it is not outrageous to extract reference values
using the Gaussian state,
which exhibits the same single-particle correlation functions
of the (non-Gaussian) time-evolved state for a finite time and finite sizes.
At least for the MI quench, as we see below,
single-particle correlations are independent of time and size.
As a result, the R\'{e}nyi entanglement entropy of
Eq.~(\ref{eq:gaussian_Renyi}) obtained from the mode occupation numbers
$n_{\mu}$ for any $tJ, L<\infty$ gives the true R\'{e}nyi entanglement entropy for $L \gg tJ \gg 1$.
Hereafter we use the symbol $S^{\mathrm{Gaussian}}_2$
to denote the R\'{e}nyi entanglement entropy
estimated from Eq.~(\ref{eq:gaussian_Renyi}).

For the quench starting from the MI state,
the matrix consisting of the correlation function is already diagonal;
$\langle \hat{b}^{\dagger}_j \hat{b}_l \rangle = \delta_{j,l}$
and $\langle \hat{b}_j \hat{b}_l \rangle=0$ $(j, l = 1, 2, \dots, L_{\mathrm{A}})$ 
for all $L$ and $t J$.
The mode occupation numbers,
$n_{\mu}=1$ ($\mu=1,2,\dots,L_{\mathrm{A}}$),
are independent of time and sizes.
Therefore,
the R\'{e}nyi entanglement entropy of subsystem A, whose size is $L_{\mathrm{A}}=L/2$, is given by
$S^{\mathrm{Gaussian}}_2=\ln 3\times L/2 \approx 0.5493\times L$.
Indeed, numerically obtained $S_2$ for $tJ \gg L$
fluctuates around $S^{\mathrm{Gaussian}}_2$
[see Figs.~\ref{fig:s2_div_L_bf}(a) and \ref{fig:s2_cmp_gaussian}(a)],
while a $t$-linear growth is not reproduced in $S^{\mathrm{Gaussian}}_2$
for a short time ($tJ\ll L$).

For the quench starting from the CDW state,
we numerically evaluate
the R\'{e}nyi entanglement entropy for a subsystem size $L_{\mathrm{A}}=L/2$.
In this case,
$\langle \hat{b}^{\dagger}_j \hat{b}_l \rangle$ depends on time,
and therefore, $S^{\mathrm{Gaussian}}_2$ oscillates in time.
Again, numerically obtained $S_2$ for $tJ\gg L$
fluctuates around $S^{\mathrm{Gaussian}}_2$
[see Figs.~\ref{fig:s2_div_L_bf}(b) and \ref{fig:s2_cmp_gaussian}(b)],
whereas a $t$-linear growth in a short time is absent for $S^{\mathrm{Gaussian}}_2$.
The R\'{e}nyi entanglement entropy in the thermodynamic limit for a long time
is estimated to be $S^{\mathrm{Gaussian}}_2/L = 0.31(1)$ for the Gaussian state,
and $S_2$ is expected to converge to this value for $L \gg tJ \gg 1$.

\subsection{Comparison with Page value}

The Bose-Hubbard model is nonintegrable (integrable) for $|U|>0$ ($U=0$).
The system is thermalized when it is quenched to the parameter region
$|U|>0$~\cite{biroli2010,sorg2014}.
The entanglement entropy would be nearly saturated 
at that of the random state vector,
which is known as the Page value $S_{2}^{\mathrm{Page}}$~\cite{page1993}.
For the Bose-Hubbard model, an analytical expression for the Page value has not been obtained yet.
We can obtain the Page value within the statistical error bars by numerically taking an average of random state vectors~\cite{russomanno2020,kunimi2021}.
When taking 
this average, we should directly use the Hilbert space of the
Bose-Hubbard model.
The dimension of the Hilbert space 
grows 
exponentially 
when increasing the system size.
Therefore, we can estimate the Page value only for rather small system sizes.
In this paper, we calculate $S_{2}^{\mathrm{Page}}$ up to $L = 12$ $(L=16)$ at unit filling (half filling).

For the $U=0$ quench, the time-evolved state is not thermalized.
The R\'{e}nyi entanglement entropy, in this case, would deviate from the Page value.
To examine whether we can tell the difference
between the entanglement entropy of the thermalized state
and that of the state quenched to $U=0$,
we compare the R\'{e}nyi entanglement entropy at $U=0$ with the Page value.
We calculate the Page value by averaging
$1024$ random samples.

\begin{figure}[!t]
\centering
\includegraphics[width=\columnwidth]{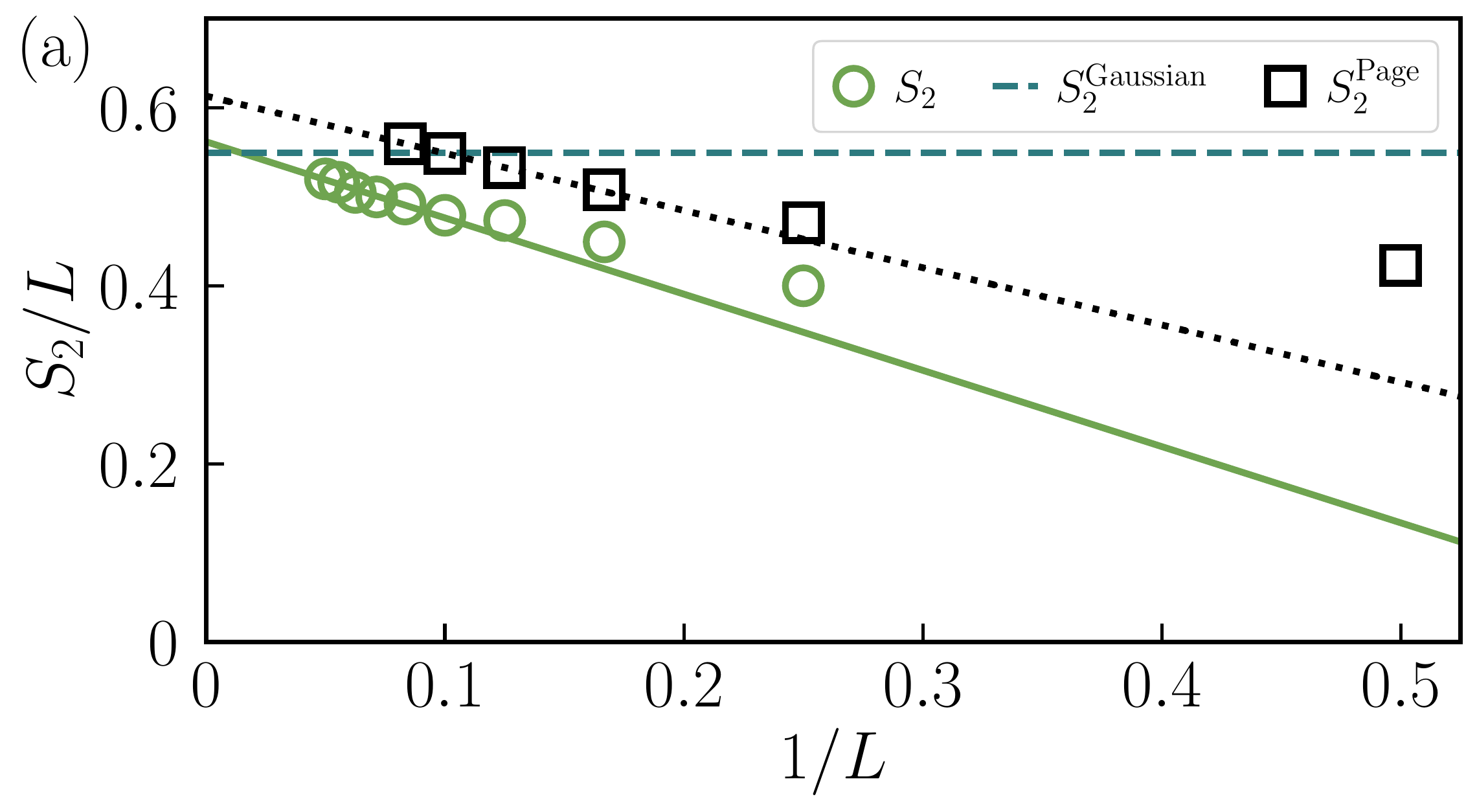}
\\
\includegraphics[width=\columnwidth]{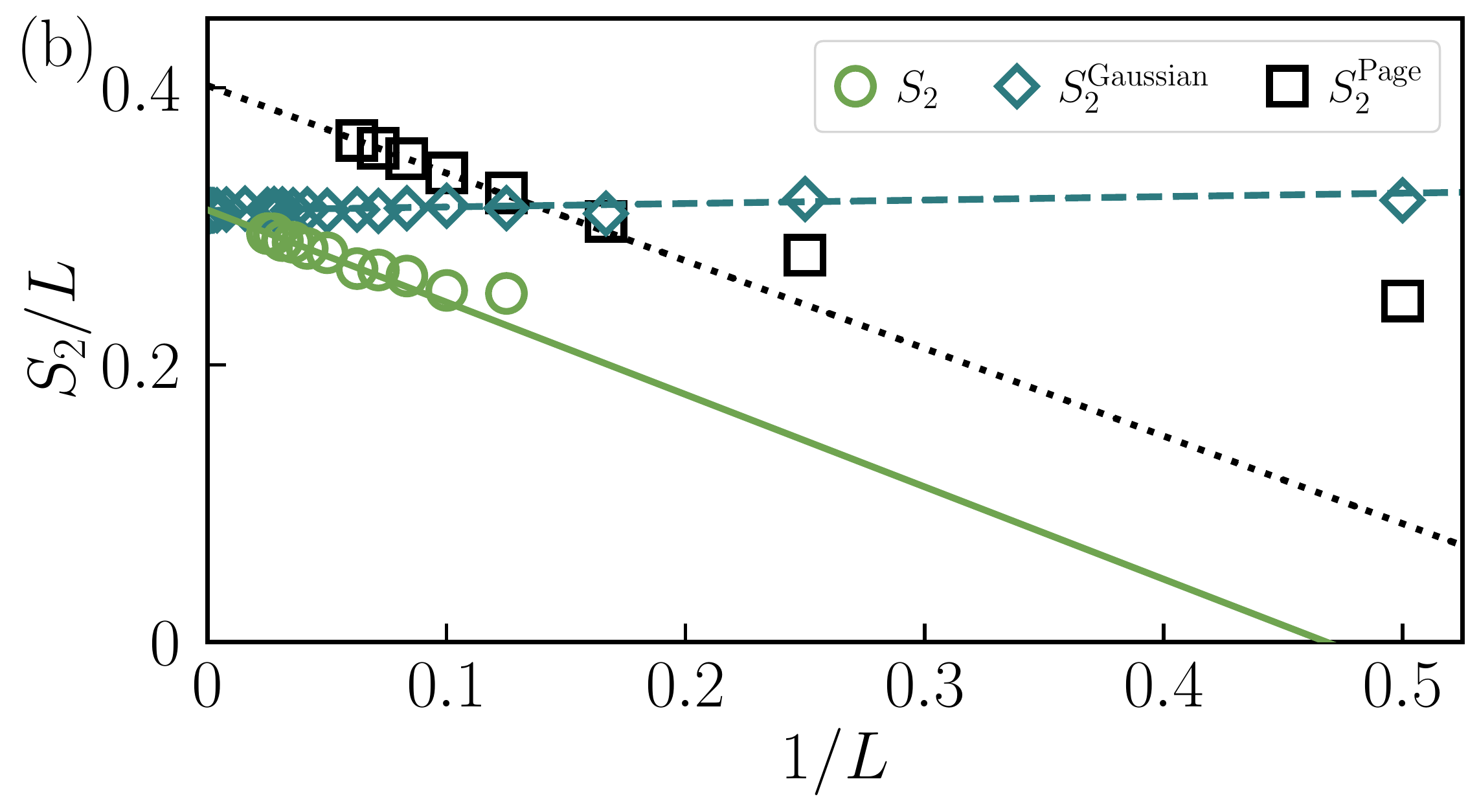}
\caption{%
System size dependence of the R\'{e}nyi entanglement entropy density for (a) the MI initial state and (b) the CDW initial state.
Circles represent the time average of the R\'{e}nyi entanglement entropy density of the time-evolved state $S_2 / L$ from $tJ = 1$ to $tJ = 10^4$.
Diamonds are the time average of $S_{2}^{\mathrm{Gaussian}} / L$ from $tJ = 1$ to $tJ = 10^4$.
Squares are the Page value.
The solid and dotted lines are, respectively, linear fits
of $S_{2} / L$ and $S_{2}^{\mathrm{Page}}/L$
using five largest systems.
The dashed line in (a) represents $S_{2}^{\mathrm{Gaussian}} / L$, which is independent of the system size $L$.
The dashed line in (b) shows a linear fit of $S_{2}^{\mathrm{Gaussian}} / L$ using data satisfying $L \geq 20$.
We do not show error bars because they are small compared to symbol sizes.
}
\label{fig:s2_size_scaling}
\end{figure}

As for the quench from the MI state,
the Page value $S^{\mathrm{Page}}_2$
at unit filling is close to the R\'{e}nyi entanglement entropy $S_2$
for a longer time $(tJ \gtrsim L/4)$
and for all systems ($L\le 12$) that we have considered
[see Fig.~\ref{fig:s2_cmp_gaussian}(a)].
When increasing the system size $L$, we see that the Page value becomes greater than the R\'{e}nyi entanglement entropy $S_2$ at a longer time.
This observation indicates that when taking the thermodynamic limit, the Page value converges to the value larger than $S_2$.
We examine the system-size dependence and extrapolate these values, as well as $S^{\mathrm{Gaussian}}_{2}$, to the thermodynamic limit, as shown in Fig.~\ref{fig:s2_size_scaling}(a).
The R\'{e}nyi entanglement entropy $S_2$ in the thermodynamic limit is very close to that of the Gaussian state, which is expected
from previous studies~\cite{cramer2008a,flesch2008,cramer2008b},
although we see a small deviation due to finite-size effects.
On the other hand, 
in the thermodynamic limit,
the Page value 
is greater than $S_2$ and $S^{\mathrm{Gaussian}}_{2}$.
Thus, it is expected that a jump occurs between the entropies of $U=0$ and $|U|>0$ quenches after long-time evolution for sufficiently large system sizes, although
they take similar values in a small system.
In this respect, we compare the R\'{e}nyi entanglement entropy of $U=0$ with that of $U > 0$ and confirm the presence of a jump in Appendix~\ref{sec:finiteUquench}.

For the quench from the CDW state
at half filling, the R\'{e}nyi entanglement entropy $S_2$ is slightly smaller than
the Page value unlike the MI case
[see Fig.~\ref{fig:s2_cmp_gaussian}(b)].
The deviation seems to be more enhanced with increasing system sizes.
It is likely that
the R\'{e}nyi entanglement entropy density would be smaller than
the density of the Page value in the thermodynamic limit
although the R\'{e}nyi entanglement entropy itself
satisfies the volume-law scaling,
as can be seen from Fig.~\ref{fig:s2_div_L_bf}(b).
We investigate the system-size dependence and extract the R\'{e}nyi entanglement entropy density in the thermodynamic limit, as shown in Fig.~\ref{fig:s2_size_scaling}(b).
From the same discussion as in the MI case, there is a jump between the entropies of $U=0$ and $|U|>0$ quenches after long-time evolution, which is discussed in Appendix~\ref{sec:finiteUquench}.

The time-averaged R\'{e}nyi entanglement entropy density $S_2/L$ of the $U=0$ quench is smaller than the Page value $S^{\mathrm{Page}}_{2} / L$, which is 
also the value $S_2/L$ expected for
the $|U| > 0$ quench, as shown in Fig.~\ref{fig:s2_size_scaling}.
This would be understood from the viewpoint of the number of states and the integrability.
When the state is thermalized, the entanglement entropy is saturated at the Page value and the state would be characterized by the thermal distribution.
Therefore, the entanglement entropy as well as the Page value would be identified as the thermal entropy, described by the logarithm of the number of states.
When we denote $\zeta$ as the number of states per site, the number of states in subsystem A is given by $\zeta^{L_{\mathrm{A}}}$ and the Page value $S^{\mathrm{Page}}_{2} / L$ would be approximated by $[\ln (\zeta)] / 2$ (where $L_{\mathrm{A}} = L / 2$).
On the other hand, after a long-time evolution, the state quenched to $U=0$ relaxes to a state characterized by the
generalized Gibbs ensemble (GGE), which is the Boltzmann (thermal) distribution
taking into account not only the internal energy but also a set of
conserved quantities~\cite{Rigol2007}.
In this case, the time-averaged R\'{e}nyi entanglement entropy can be
seen as the logarithm of the number of states in the GGE~\cite{Nakagawa2018}.
Due to the presence of the exponentially large number of conserved
quantities, the number of states in the GGE would be $\zeta^{L_{\mathrm{A}}} / \eta^{L_{\mathrm{A}}}$, where $\eta~(>1)$ represents the number of conserved charges per site.
Consequently, the entropy density of the GGE, $S_2 / L$ is estimated as $[\ln( \zeta / \eta)] / 2$.
Comparing the Page value and the entropy density of the GGE, we see that $S_2 / L$ is always smaller than $S_2^{\mathrm{Page}} / L$.

\subsection{%
Entropy-density-like value and practical bound for the R\'{e}nyi
entanglement entropy}
\label{subsec:relation_inf_norm}

\begin{figure}[!t]
\centering
\includegraphics[width=\columnwidth]{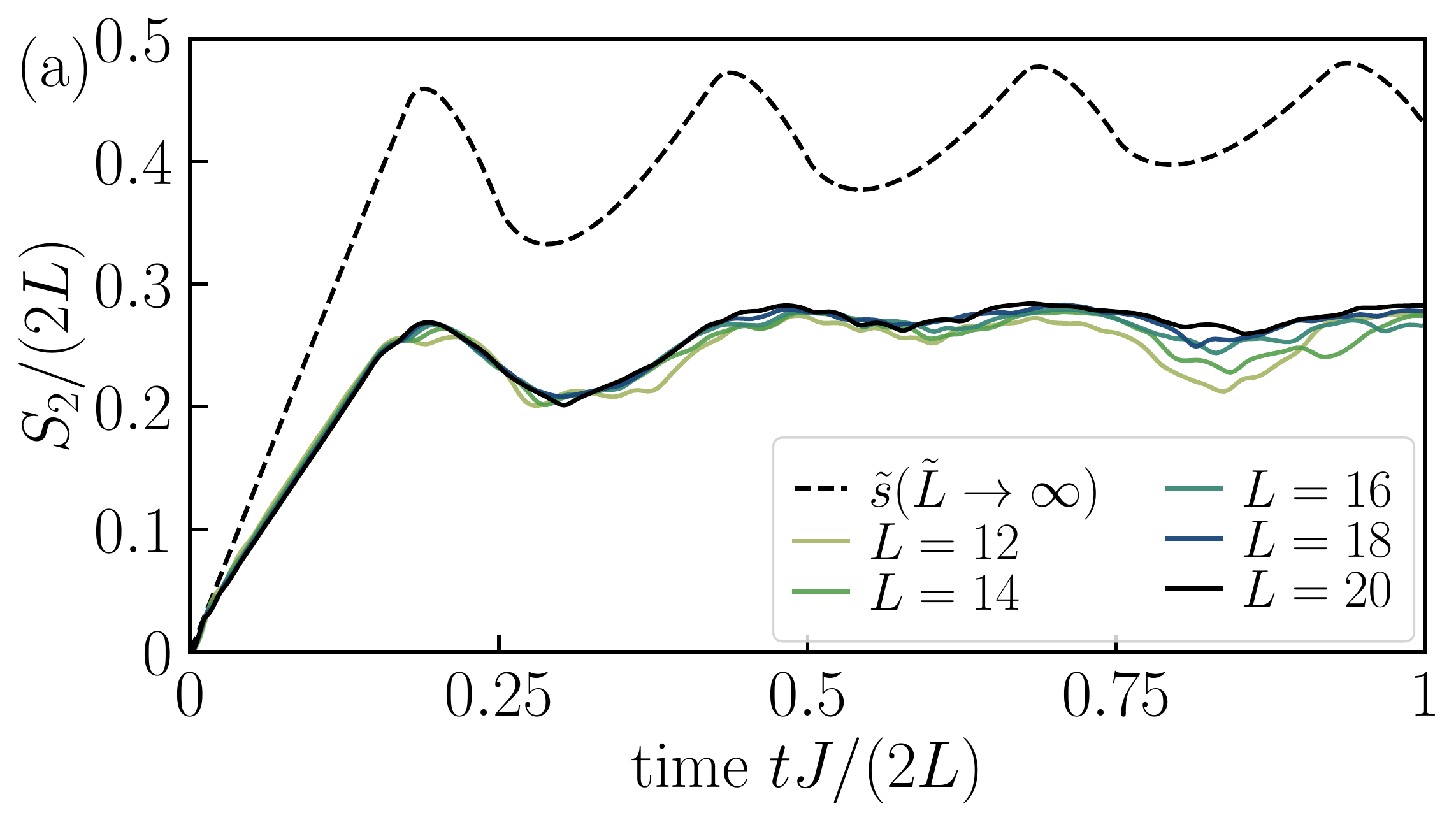}
\\
\includegraphics[width=\columnwidth]{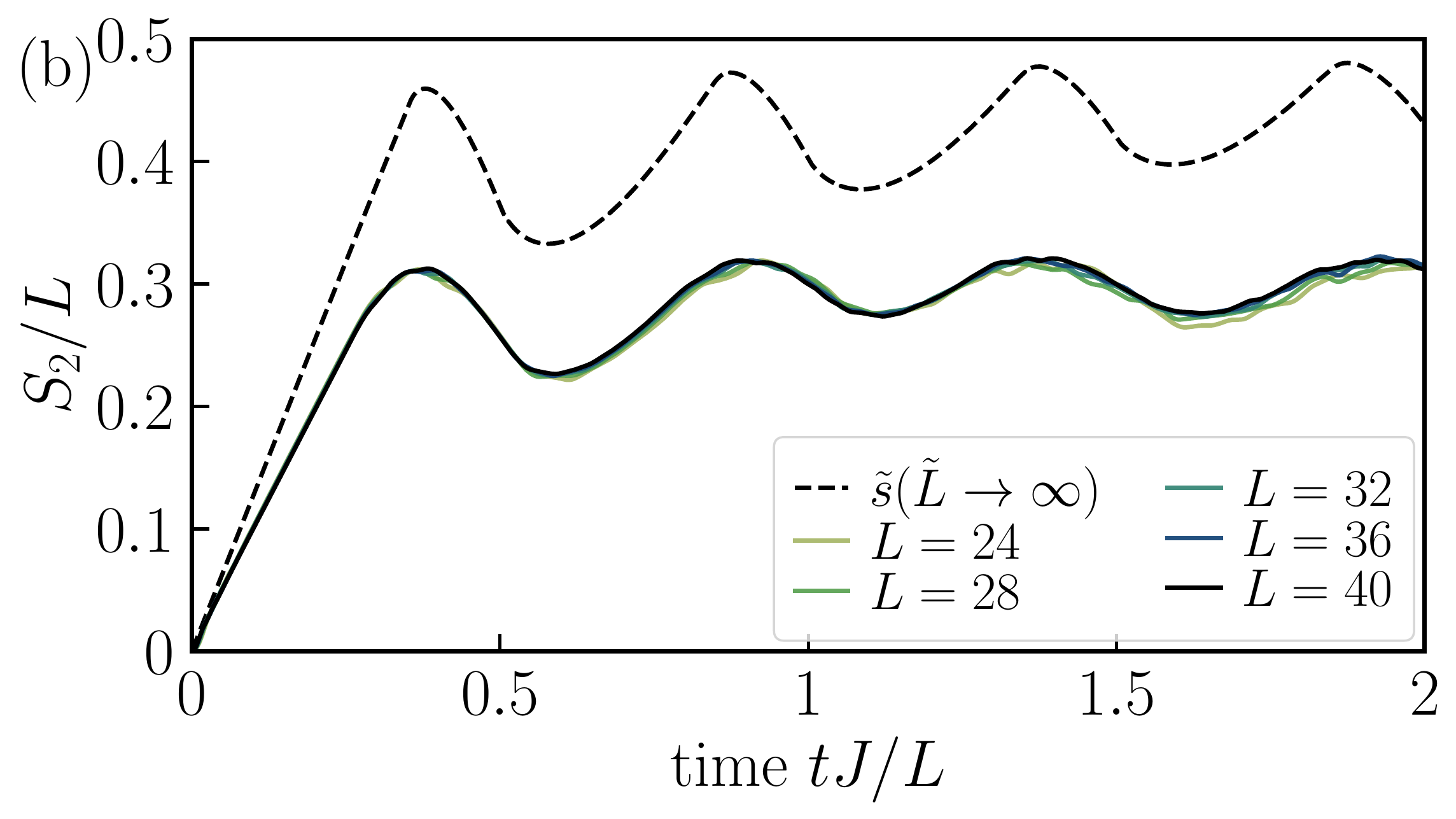}
\caption{Comparisons between
the rescaled R\'{e}nyi entanglement entropy $S_2/\tilde{L}$ and
$\tilde{s}(\tilde{L})$
estimated from the infinity norm of rows of the matrix $A_Z$
for the quench from the (a) MI ($\tilde{L}=2L$) and
(b) CDW ($\tilde{L}=L$) states.
The value $\tilde{s}(\tilde{L} \to \infty)$ is obtained for
a sufficiently large size ($L=1024$).
Note that $\tilde{s}(\tilde{L})$'s are the same for the MI and CDW states as shown in Appendix~\ref{sec:properties_tilde_s}.
}
\label{fig:s2_div_L_cmp_hinf}
\end{figure}

\begin{figure}[!t]
\centering
\includegraphics[width=\columnwidth]{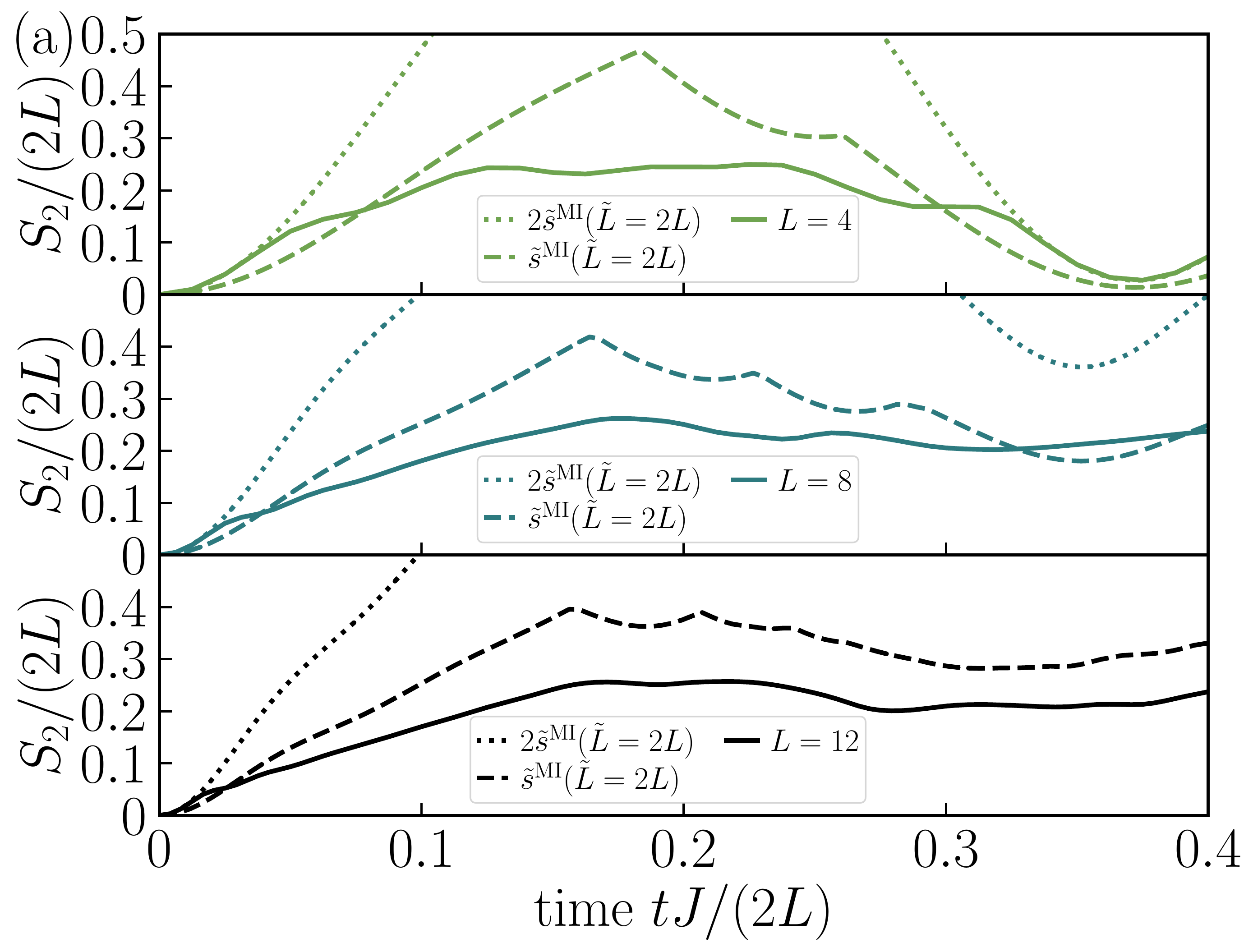}
\\
\includegraphics[width=\columnwidth]{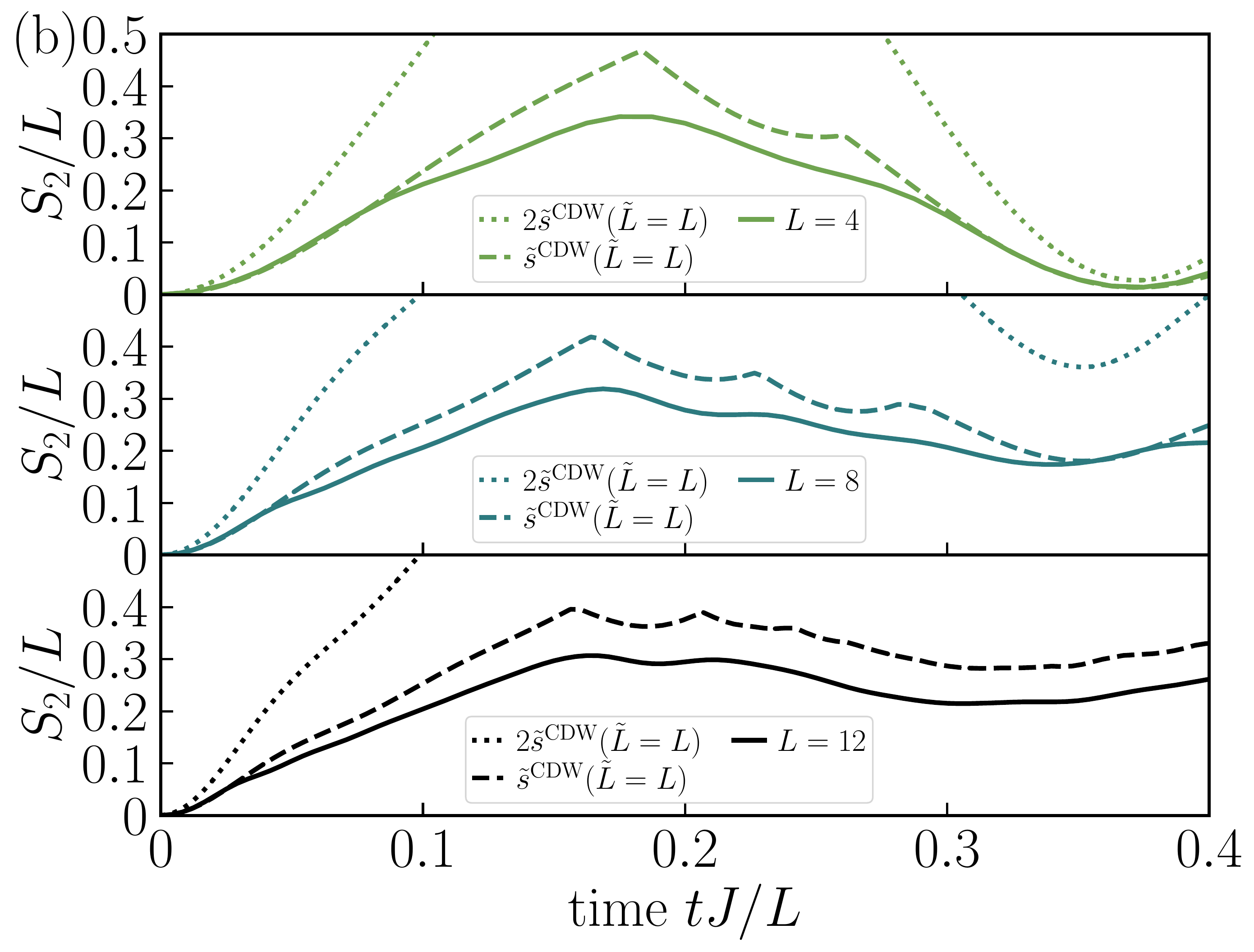}
\caption{Comparisons between
the rescaled R\'{e}nyi entanglement entropy $S_2/\tilde{L}$ and
$\tilde{s}(\tilde{L})$
estimated from the infinity norm of rows of the matrix $A_Z$
for the quench from the (a) MI ($\tilde{L}=2L$) and
(b) CDW ($\tilde{L}=L$) states
for $L=4,8,~\text{and}~12$.
In both cases, the R\'{e}nyi entanglement entropy density $S_2/L$
seems to be bounded by $2\tilde{s}(\tilde{L})$.
Note that $\tilde{s}(\tilde{L})$'s are the same
for the MI and CDW states as shown in Appendix~\ref{sec:properties_tilde_s}.
}
\label{fig:s2_div_L_cmp_hinf_smaller}
\end{figure}

From the argument in Sec.~\ref{subsec:volume_law},
we obtain the rigorous lower bound for the second
R\'{e}nyi entanglement entropy density, which is given as
\begin{align}
 \frac{S_2}{\tilde{L}}
 \ge 10^{-5} \times \left[ 1-g_{A_Z}(\tilde{L}) \right]^2.
\end{align}
Moreover, the tighter bound was conjectured to be
$\mathrm{perm} A\le \exp \{-\mathrm{const} \times M [1 - g_{A}(M)] \},
$
where $M$ is the size of the matrix $A$~\cite{berkowitz2018}, which gives the lower bound of the
R\'{e}nyi entanglement entropy as
\begin{align}
\frac{S_2}{\tilde{L}} \geq \mathrm{const} \times \left[1 - g_{A_Z} (\tilde{L}) \right].
\label{eq:lowerbound}
\end{align}
We briefly note that the conjectured bound gives the same condition for
the volume-law entanglement growth as Eq.~(\ref{eq:condition_volume_law}).
These inequalities lead us to expect that 
$\tilde{s}(\tilde{L}) = 1 - g_{A_Z}(\tilde{L})$
determines the qualitative behavior of the R\'{e}nyi
entanglement entropy density
and serves as an entropy-density-like value.
$\tilde{s}(\tilde{L})$
is obtained from $g_{A_Z}$ in Eq.~(\ref{eq:g_A_Z_definition}):
\begin{align}
  \tilde{s}(\tilde{L})
  = 1 - \frac{1}{N} \sum_{j=1}^{N} \max_{l=1}^{N} ( |z_{j,l}|,|\delta_{j,l} - z_{j,l}| ).
  \label{eq:stilde_definition}
\end{align}
As shown in Appendix~\ref{sec:properties_tilde_s}, the entropy-density-like value can be expressed in a simple form:
\begin{align}
\tilde{s}(\tilde{L})
=
\frac{1}{2} - \frac{1}{N} \sum_{j=1}^{N} \left| z_{j,j} - \frac{1}{2} \right|.
\label{eq:stilde_simple_form}
\end{align}
Unlike the permanent of matrix $A_Z$, which requires costly calculation, 
$\tilde{s}(\tilde{L})$
 is easy to compute numerically and analytically.
Hence if 
$\tilde{s}(\tilde{L})$
has a similar tendency with 
$S_2 / \tilde{L}$
, it would be a helpful quantity to qualitatively capture features of the entanglement entropy density.

To see the behavior of 
$\tilde{s}(\tilde{L})$
in the thermodynamic limit, we compare
$\tilde{s}(\tilde{L})$
for a much larger system $L=1024$
with $S_2/\tilde{L}$ for $\tilde{L} \leq 40$
in Fig.~\ref{fig:s2_div_L_cmp_hinf}.
We have confirmed that $\tilde{s}(\tilde{L})$ 
converges 
well 
when $L \gtrsim 256$
and, therefore,
regarded $\tilde{s}(\tilde{L})$ with $L = 1024$ as that in the thermodynamic limit,
$\tilde{s}(\tilde{L} \to \infty)$.
Both $\tilde{s}(\tilde{L} \to \infty)$ and $S_2/\tilde{L}$
are found to behave in a qualitatively similar way
including the period of oscillations.
Thus, we confirm that $\tilde{s}(\tilde{L} \to \infty)$
certainly captures the
qualitative behavior of the R\'{e}nyi entanglement entropy density.

While Eq.~(\ref{eq:lowerbound}) means that ${\rm const}\times
\tilde{s}(\tilde{L})$ serves as a lower bound of $S_2/\tilde{L}$, we
observe in Fig.~\ref{fig:s2_div_L_cmp_hinf} that the rescaled R\'{e}nyi
entanglement entropy $S_2/\tilde{L}$
would be practically bounded from above
by the entropy-density-like value in the thermodynamic limit $\tilde{s}(\tilde{L} \to \infty)$.
This observation leads us to expect that $\tilde{s}(\tilde{L})$ would 
play a role of an upper bound of $S_2 / \tilde{L}$ for any system sizes and 
motivates us to examine how $\tilde{s}(\tilde{L})$
bounds the R\'{e}nyi entanglement entropy $S_2 / \tilde{L}$
from above.
For this purpose,
we compare these for
some finite $L$'s,
as shown in Fig.~\ref{fig:s2_div_L_cmp_hinf_smaller}.
We find a region where the expected inequality
$S_2 / \tilde{L} < \tilde{s}(\tilde{L})$ is slightly violated.
Even in this case, for the MI quench,
$S_2/\tilde{L}$ is bounded by $2\tilde{s}(\tilde{L})$ from above and they well overlap in the time range $tJ/(2L)\lesssim 0.02$ [see Fig.~\ref{fig:s2_div_L_cmp_hinf_smaller}(a)].
Likewise, in the case of the CDW quench,
$2\tilde{s}(\tilde{L})$ still gives an upper bound
for $S_2/\tilde{L}$
[see Fig.~\ref{fig:s2_div_L_cmp_hinf_smaller}(b)].
From these observations, we speculate that some constant value times
$\tilde{s}(\tilde{L})$ would practically give an upper bound of the
R\'{e}nyi entanglement entropy.

To sum up this section, we have observed
\begin{align}
 \frac{S_2}{\tilde{L}}
 &\lesssim  \tilde{s}(\tilde{L} \to \infty)
 \quad \text{for $\tilde{L}\gg 1$},
 \\
 \frac{S_2}{\tilde{L}}
 &\lesssim  \mathrm{const.} \times \tilde{s}(\tilde{L})
 \quad \text{for finite $\tilde{L}$}.
\end{align}
These results imply that $\tilde{s}(\tilde{L})$, which can be obtained from the infinity norm of rows of the matrix $A_Z$,
would be a helpful guide in qualitatively estimating the R\'{e}nyi entanglement entropy at least in the present case.
We expect that the current discussion is applicable to other initial conditions and Hamiltonians.

\section{Summary and outlook}
\label{sec:summary}

We have studied the time evolution of the R\'{e}nyi entanglement entropy
in a 1D free boson system.
We have focused on the quench dynamics of the 1D Bose-Hubbard model at the noninteracting point starting from the Mott-insulating and charge-density-wave initial states.
We have obtained the analytical form of the second R\'{e}nyi
entanglement entropy
by calculating the expectation value of the shift operator.
The R\'{e}nyi entanglement entropy was found to be the negative of the logarithm of the permanent of the matrix
whose elements are time-dependent single-particle correlation functions.
Using a permanent inequality,
we have rigorously proven that the R\'{e}nyi entanglement entropy satisfies
the volume-law scaling under a certain condition.
We have also numerically obtained the long-time evolution of
the R\'{e}nyi entanglement entropy by direct computations of the permanent.
Although it requires exponential time in general,
the present approach is superior to the best currently available methods
such as the exact diagonalization and matrix-product-state methods.
The feasible system size is about twice the size
that the conventional method can handle~\cite{goto2019}.
Since our method enables us to compute the R\'{e}nyi entanglement entropy at any time, the reachable time is also much
longer than the conventional methods.

The procedure presented in this paper can be extended
to systems containing long-range hopping parameters 
and those with randomness.
Real-time dynamics of such complex quantum many-body systems
of free fermions have attracted
much attention recently, whereas those of free bosons
are yet to be explored because of their computational difficulties.
Our method would be helpful for studying such bosonic counterparts.
Typical examples include
(i) noninteracting higher-dimensional systems
(see Refs.~[\onlinecite{carleo2014,nagao2019,kaneko2022}]
for correlation-spreading dynamics with an interaction quench in two
dimensions),
(ii) Anderson localization with long-range hopping
(see Refs.~[\onlinecite{modak2020}] and
[\onlinecite{singh2017}] for free fermions),
(iii) localization in disorder-free or correlated-disorder systems
such as the Aubry-Andr\'{e} model~\cite{aubry1980}
(see Refs.~[\onlinecite{modak2020}] and
[\onlinecite{devakul2017}] for free fermions), and
(iv) Lindblad dynamics of free bosons
(see Refs.~[\onlinecite{cao2019,alberton2021,minato2022,block2022,muller2022}]
for free fermions).
Our formula would also be useful in studying the entanglement properties of mixtures of bosons and fermions (e.g., fermion system in cavities~\cite{Cordoba2020} and Bose-Fermi-Hubbard systems~\cite{Arenas2021}).

We have introduced the entropy-density-like value using
the infinity norm of rows of the matrix consisting of the correlation function and
have numerically demonstrated that it well captures the features of the
R\'{e}nyi entanglement entropy density.
We have also discussed a practical bound of the entanglement entropy, i.e., that of the matrix permanent,
which usually requires exponential time computations.
Our findings on the practical bound
would also stimulate mathematical research on the permanent of
general complex matrices
and research in the field of quantum computing
involving boson-sampling techniques~\cite{aaronson2011,aaronson2014}.

\acknowledgments

The authors acknowledge fruitful discussions with
S.\ Goto and Y.\ Takeuchi.
This work was financially supported by JSPS KAKENHI
(Grants No.\ 18H05228, No.\ 19K14616, No.\ 20H01838, No.\ 21H01014, and
No.\ 21K13855),
by Grant-in-Aid for JSPS Fellows (Grant No.\ 22J22306),
by JST CREST (Grant No.\ JPMJCR1673),
by MEXT Q-LEAP (Grant No.\ JPMXS0118069021),
and by JST FOREST (Grant No.\ JPMJFR202T).

\appendix

\section{Derivation of equations (\ref{eq:action_shift_operator_1}) and
(\ref{eq:action_shift_operator_2})}
\label{sec:action_shift_operator}

Although we apply Eqs.~(\ref{eq:action_shift_operator_1}) and
(\ref{eq:action_shift_operator_2}) to the product state of the same state living in copies 1 and 2, such as Eq.~(\ref{eq:total_wave_func}), in the main part, it holds even if states of copies 1 and 2 are different.
To prove this, we first Schmidt decompose the state of copy 1(2) as
\begin{align}
|\psi \rangle^{1(2)}
=
\sum_{l} s_l^{1(2)} |\phi_{l} \rangle^{1(2),\mathrm{A}} |\varphi_{l} \rangle^{1(2),\mathrm{B}},
\end{align}
where $|\phi_{l} \rangle^{1(2),\mathrm{A}}$ and $|\varphi_{l} \rangle^{1(2),\mathrm{B}}$ are orthonormal states of subsystems A and B, respectively.
$s_l^{1(2)}$ is the Schmidt coefficient.
The product state of copies 1 and 2 is given by
\begin{align}
|\psi_{\mathrm{prod}} \rangle = |\psi \rangle^{1} \otimes |\psi' \rangle^{2},
\label{eq:prod_wave_func}
\end{align}
where we assume that copies 1 and 2 can have different states.

We consider how $\hat{V}_{\mathrm{A}} \hat{b}^{\dag}_{j} \hat{V}_{\mathrm{A}}^{-1}$ acts on the state given by Eq.~(\ref{eq:prod_wave_func}).
Note that
$\hat{V}_{\mathrm{A}}^{-1} = \hat{V}_{\mathrm{A}}$
holds because of $\hat{V}_{\mathrm{A}}^2 = \hat{I}$, where $\hat{I}$ is an identity operator.
When $j \in \mathrm{B}$, $\hat{V}_{\mathrm{A}} \hat{b}^{\dag}_{j} \hat{V}_{\mathrm{A}}^{-1} |\psi_{\mathrm{prod}} \rangle = \hat{b}^{\dag}_{j} |\psi_{\mathrm{prod}} \rangle$ trivially holds.
In the case of $j \in \mathrm{A}$, the action of $\hat{V}_{\mathrm{A}} \hat{b}^{\dag}_{j} \hat{V}_{\mathrm{A}}^{-1}$ on the product state is given by
\begin{align}
 &~\phantom{=}~   
 \hat{V}_{\mathrm{A}} \hat{b}_j^{\dag} \hat{V}_{\mathrm{A}}^{-1}
|\psi_{\mathrm{prod}} \rangle
 \nonumber
\\
 &= \hat{V}_{\mathrm{A}} \hat{b}_j^{\dag} \hat{V}_{\mathrm{A}}^{-1}
\sum_{l,m} s^{1}_l s'^{2}_m |\phi_{l} \rangle^{1,\mathrm{A}} |\varphi_{l}
\rangle^{1,\mathrm{B}}
 |\phi'_{m} \rangle^{2,\mathrm{A}} |\varphi'_{m} \rangle^{2,\mathrm{B}}
 \nonumber
\\
 &=
 \hat{V}_{\mathrm{A}} \hat{b}_j^{\dag} \sum_{l,m} s^{1}_l s'^{2}_m
|\phi'_{m} \rangle^{1,\mathrm{A}} |\varphi_{l} \rangle^{1,\mathrm{B}}
 |\phi_{l} \rangle^{2,\mathrm{A}} |\varphi'_{m} \rangle^{2,\mathrm{B}}
 \nonumber
\\
&=
\hat{V}_{\mathrm{A}} \sum_{l,m} s^{1}_l s'^{2}_m (\hat{b}^{\dag}_{j}
|\phi'_{m} \rangle^{1,\mathrm{A}}) |\varphi_{l} \rangle^{1,\mathrm{B}}
 |\phi_{l} \rangle^{2,\mathrm{A}} |\varphi'_{m} \rangle^{2,\mathrm{B}}
 \nonumber
\\
 &=
 \sum_{l,m} s^{1}_l s'^{2}_m |\phi_{l} \rangle^{1,\mathrm{A}}
|\varphi_{l} \rangle^{1,\mathrm{B}}
 (\hat{c}^{\dag}_{j} |\phi'_{m} \rangle^{2,\mathrm{A}})|\varphi'_{m}
\rangle^{2,\mathrm{B}}
 \nonumber
\\
 &=
 \hat{c}_{j}^{\dag} |\psi_{\mathrm{prod}}\rangle.
\end{align}
Thus, we proved Eq.~(\ref{eq:action_shift_operator_1})
[and Eq.~(\ref{eq:action_shift_operator_2}) in the same manner]
as long as it acts on the product state of copies 1 and 2.

\section{More on the property of the matrix \texorpdfstring{$Z^{\mathrm{MI}}$}{Z}}
\label{sec:more_matrix_z}

The matrix $X$, having the element $x_{k,l}$ in Eq.~(\ref{eq:X_definition}), 
is unitary and diagonalizes
the single-particle Hamiltonian matrix $H$,
namely,
$HX=XE$, with the matrix $E$ being a diagonal matrix
consisting of all the corresponding eigenvalues.
The matrix $D=e^{-iEt}$ $(t\ge 0)$ is also unitary,
and the matrix $Y$, having the component $y_{j,l}$ in Eq.~(\ref{eq:Y_definition}),
is given as $Y=X^{\dagger}DX$.
Because $YY^{\dagger}=Y^{\dagger}Y=I$, $Y$ is also unitary.

The Hermitian matrix $Z^{\mathrm{MI}}$
can be represented as
\begin{align}
 Z^{\mathrm{MI}} = (Y P_{L/2} Y^{\dag})^T,
 \quad
 P_{L/2} = \begin{pmatrix}
 I_{L/2} & 0_{L/2} \\ 0_{L/2} & 0_{L/2}
 \end{pmatrix},
\label{eq:z_MI_diagonalization}
\end{align}
where $I_{L/2}$ and $0_{L/2}$ are, respectively, an $L/2\times L/2$ identity matrix and an $L/2\times L/2$ zero matrix.
Since $P_{L/2}$ is a projection matrix, satisfying $P_{L/2}^2 = P_{L/2}$, we immediately obtain $(Z^{\mathrm{MI}})^2 = Z^{\mathrm{MI}}$.
Therefore, the matrix $Z^{\mathrm{MI}}$ is a projection matrix.
Note that this argument holds for
any Hermitian single-particle Hamiltonian matrix $H$,
in particular, containing
long-range hopping parameters with randomness.

\section{Derivation of equation~(\ref{eq:z_approximation})}
\label{sec:breaking_volume_law}
The purpose of this appendix is to prove Eq.~(\ref{eq:z_approximation}) for a time-evolved state with a short time, $t \ll L / v_{\mathrm{C}}$.
Corresponding to the limit in Eq.~(\ref{eq:condition_volume_law}), 
we consider the thermodynamic limit, $L \to \infty$ with a fixed time $t$
in the following.

Since we consider the case $tJ \ll L$ and the thermodynamic limit, the summation in Eq.~(\ref{eq:Y_definition}) can be replaced with the integral.
In this case, the single-particle wave function $y_{j,l}$ and the correlation function $z_{j,l}$ can be expressed as
\begin{align}
y_{j,l} &= (-i)^{j-l} \left[ J_{j-l} - (-1)^{l} J_{j+l} \right],
\\
\nonumber
z_{j,l} &= i^{j} (-i)^{l}  \sum_{m=1}^{L_{\mathrm{A}}} \Big[ J_{j-m} J_{l-m} + J_{j+m} J_{l+m}
\\
&~\phantom{=}~
 - (-1)^{m} J_{j-m} J_{l+m} - (-1)^{m} J_{j+m} J_{l-m} \Big],
\label{eq:z_thermodynamic_limit}
\end{align}
where we use an abbreviation $J_{n} = J_{n}(2tJ)$ with $J_{n}(x)$ being the Bessel function of the first kind.

We note that the Bessel function is exponentially small when $|n| > (e/2) x$ because of the inequality of the Bessel function,
\begin{align}
|J_{n}(x)| \leq \frac{(x/2)^{|n|}}{|n|!} \approx \frac{e^{-(|n| + 1/2) \ln(2|n|/ex)}}{\sqrt{\pi e x}},
\label{eq:Bessel_func_bound}
\end{align}
where
$x\ge 0$
and $n$ is an integer.
Note that $J_{-n}(x) = (-1)^n J_{n}(x)$ holds.
When deriving the right-hand side of Eq.~(\ref{eq:Bessel_func_bound}),
we assume that $|n|$ is large and use Stirling's approximation.
Using this fact, we can simplify the correlation function $z_{j,l}$.
Let $\beta$ be the smallest positive integer for which $|J_{\beta}(2tJ)|$ is negligibly small.
When we 
specifically
demand $|J_{\beta}(2tJ)| < \varepsilon$, $\beta$ is approximately given by
\begin{align}
\beta \gtrsim \frac{e}{2} (2tJ) - \ln \sqrt{\pi e (2tJ)} \varepsilon + \cdots,
\end{align}
where we assume that $2tJ \gg 1$ and $\beta \gg 1$.
By definition, $J_{n}(2tJ)$ for $|n| > \beta$ is exponentially
small.

We
evaluate $z_{j,l}$ in detail using the
aforementioned properties
of the Bessel function.
When (i) $j > L_{\mathrm{A}} + \beta$, (ii) $j < L_{\mathrm{A}} - \beta$,
(iii) $l > L_{\mathrm{A}} + \beta$,
and (iv) $l < L_{\mathrm{A}} - \beta$,
we can approximate $z_{j,l}$ to a simple form.
When (i) $j > L_{\mathrm{A}} + \beta$, all terms in the summation of
Eq.~(\ref{eq:z_thermodynamic_limit}) are exponentially small because
the smallest index ($j-m$ or $j+m$) of the Bessel functions
with varying $m$ is $j - L_{\mathrm{A}} (> \beta)$.
Thus, we can approximate the correlation function as $z_{j,l} \approx 0$.
A similar argument holds for case (iii)
by replacing $j$ with $l$.

Before moving to the next case, we note that due to the unitarity of
$y_{j,l}$, the correlation function
$z_{j,l}$ can be regarded by $\delta_{j,l}$ when the upper limit of the summation
($L_{\mathrm{A}}$) in Eq.~(\ref{eq:z_thermodynamic_limit})
can be
replaced by
$L$.
This can be approximately achieved when
$\sum_{m=L_{\mathrm{A}}+1}^{L} y_{j,m}^* y_{l,m}$ is negligibly small.
[Compare the definition of $z_{j,l}$
in Eq.~(\ref{eq:def_zij}).]
When (ii) $j < L_{\mathrm{A}} - \beta$, the index with the smallest absolute value of the Bessel
functions in $\sum_{m=L_{\mathrm{A}}+1}^{L} y_{j,m}^* y_{l,m}$ with
varying $m$ is $j - (L_{\mathrm{A}} + 1) (< - \beta)$ and we can approximate $\sum_{m=L_{\mathrm{A}}+1}^{L} y_{j,m}^* y_{l,m} \approx 0$.
Therefore, 
we can regard $z_{j,l}$ as
$\delta_{j,l}$.
For case (iv), we can obtain the same result
by replacing $j$ with $l$.

As a result, $Z^{\mathrm{MI}}$ has a structure shown in
Fig.~\ref{fig:abs_z_time_dependence}.
Denoting the dense matrix part 
that does not fall under
the above
simplifications as $Z'_{2\beta}$ (whose matrix size is
$2\beta \times 2\beta$), we obtain Eq.~(\ref{eq:z_approximation}).

\section{Quench to finite \texorpdfstring{$U$}{U} from the MI and CDW initial states}
\label{sec:finiteUquench}

\begin{figure}[!t]
\centering
\includegraphics[width=\columnwidth]{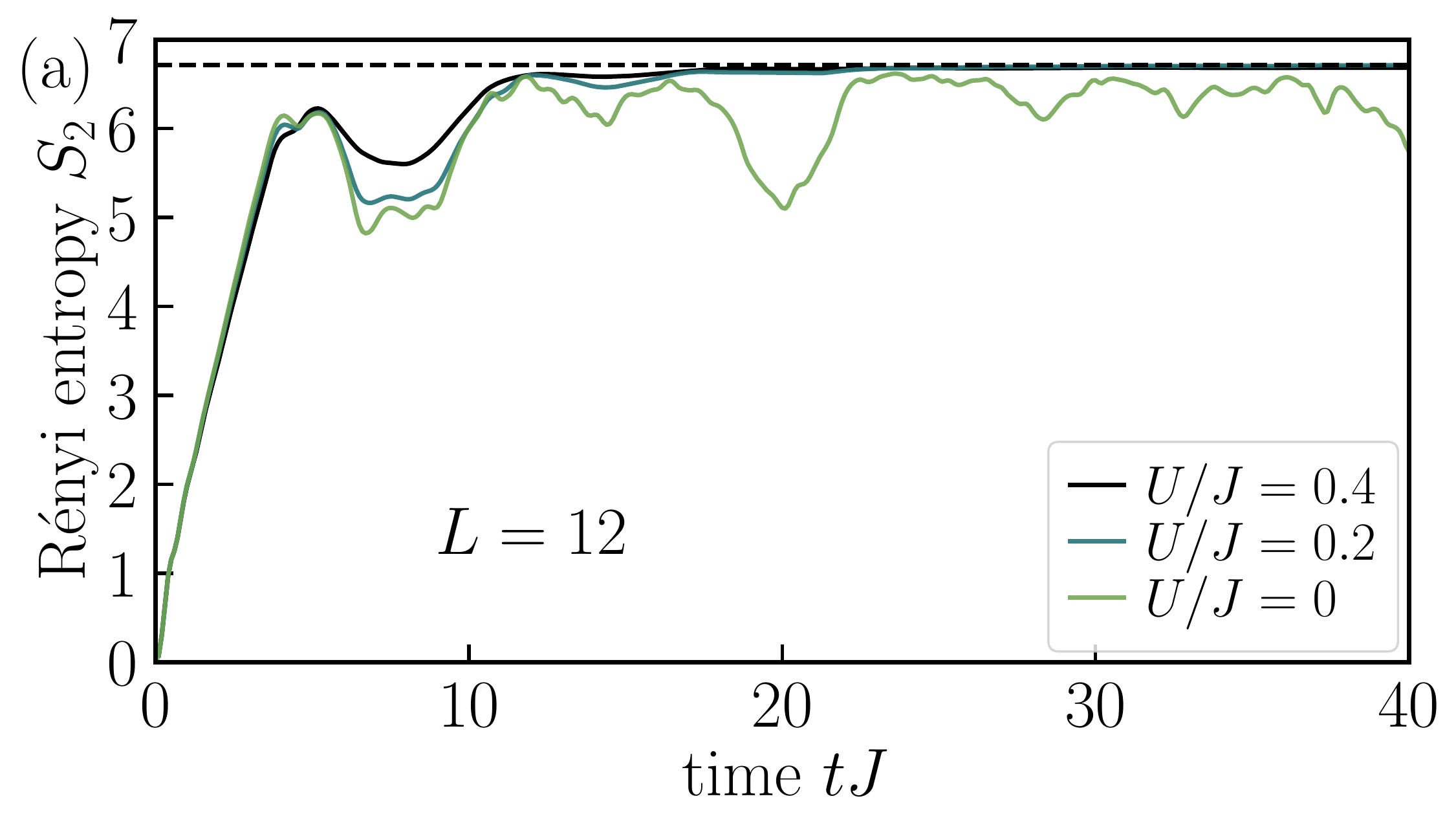}
\\
\includegraphics[width=\columnwidth]{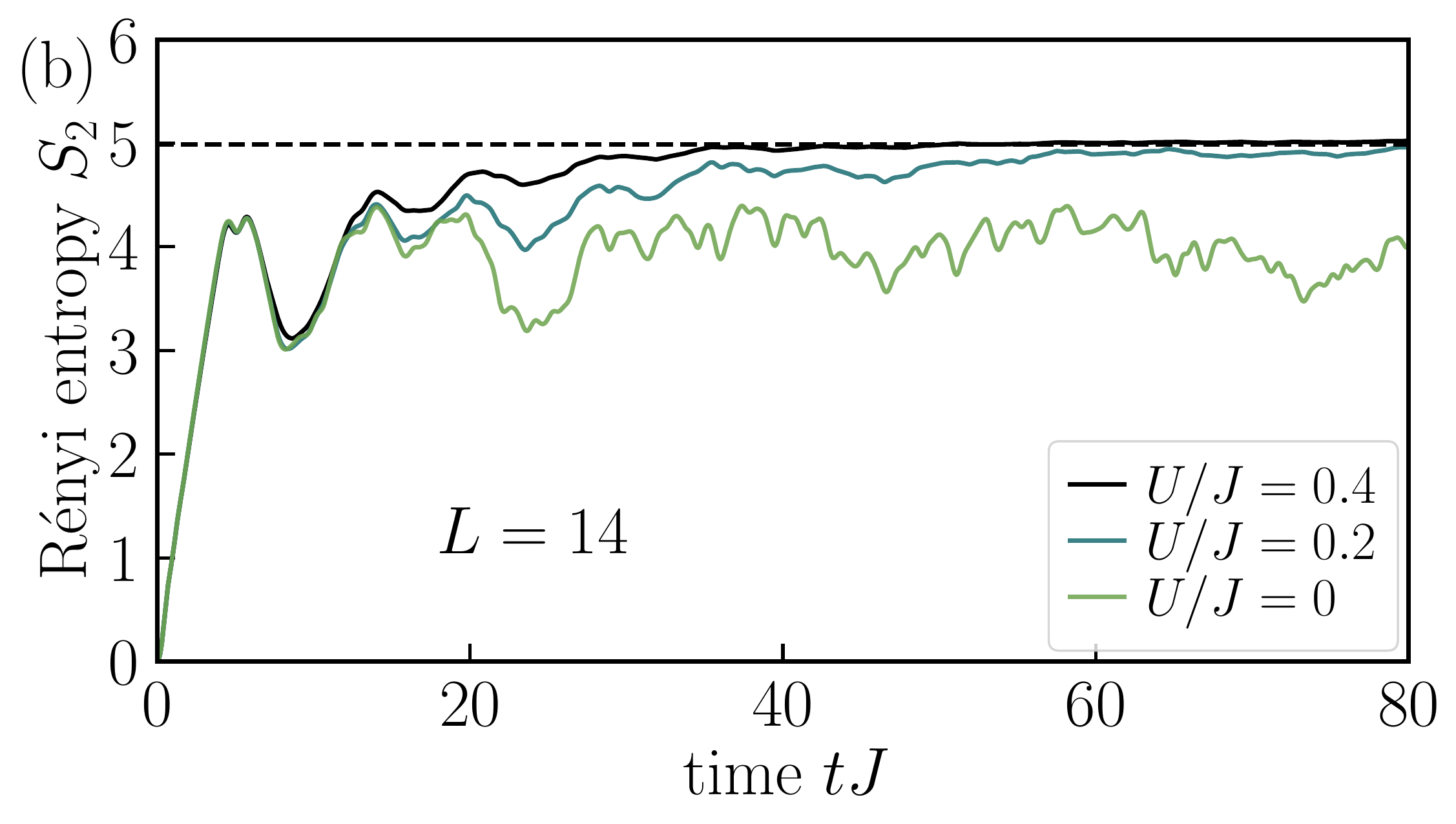}
\caption{Comparisons of the time evolution of the R\'{e}nyi entropies for
the quench to $U/J=0,~0.2$, and $0.4$
from the (a) MI and (b) CDW initial states.
The dashed line shows the Page value as the expected entanglement entropy when thermalization occurs.
For finite $U/J$, the R\'{e}nyi entanglement entropy is calculated from the exact diagonalization method~\cite{Weinberg:2017tf,Weinberg:2019vq}.
To reduce the memory cost, it is calculated
from a wave function utilizing U(1)
symmetry~\cite{jung2020},
associated with the conservation of the total particle
number~\cite{schnack2008,zhang2010,szabados2012,raventos2017}.
}
\label{fig:CDW_finiteU}
\end{figure}

As shown in Figs.~\ref{fig:s2_cmp_gaussian} and \ref{fig:s2_size_scaling}, when the system is
quenched to the noninteracting point ($U=0$) starting from the MI and CDW
initial state, the R\'{e}nyi entanglement entropy after the long-time evolution deviates from the Page value.
When the system is quenched to finite $U$, it is expected that
thermalization occurs and the R\'{e}nyi entanglement entropy approaches the Page value after the long-time evolution.
These facts indicate that there is a jump between saturated values of
the R\'{e}nyi entanglement entropy for the $U=0$ and finite-$U$ quenches.

To confirm this, we calculate the time evolution of the R\'{e}nyi
entanglement entropy for the quench to finite $U$
starting from the MI and CDW states
by the exact diagonalization method~\cite{Weinberg:2017tf,Weinberg:2019vq}
and compare it with that for the $U=0$ quench, as shown in
Figs.~\ref{fig:CDW_finiteU}(a) and (b).
In both cases, we observe the jump between the R\'{e}nyi entanglement entropies of $U=0$ and small but finite $U$.
We also find that the R\'{e}nyi entanglement entropies for finite $U$ converge to the Page value.
Thus, we can distinguish whether the state is thermalized (for $U>0$) or
not (for $U=0$) from the R\'{e}nyi entanglement entropy in a sufficiently large
system~(see Figs.~\ref{fig:s2_cmp_gaussian} and \ref{fig:s2_size_scaling}).

\section{Properties of the entropy-density-like value}
\label{sec:properties_tilde_s}
In this appendix, we show the derivation of Eq.~(\ref{eq:stilde_simple_form}) and analytically prove the agreement of the
entropy-density-like value $\tilde{s}(\tilde{L})$ of the MI initial
state and 010101$\cdots$ CDW initial state with the same system size $L$ as seen in Figs.~\ref{fig:s2_div_L_cmp_hinf} and \ref{fig:s2_div_L_cmp_hinf_smaller}.

To show Eq.~(\ref{eq:stilde_simple_form}), we first point out that there
is an upper bound of elements of the matrix $Z$.
This can be followed by the fact that $2Z^{\mathrm{MI}} - I$ is a unitary matrix.
Indeed, the relation
\begin{align}
\nonumber
&\phantom{=}~
\left( 2Z^{\mathrm{MI}} - I \right)^{\dag} \left( 2Z^{\mathrm{MI}} - I \right)
\\
&=
(Y^{\dag})^T \left( 2P_{L/2} - I \right) Y^T (Y^{\dag})^T\left( 2P_{L/2} - I \right) Y^T = I
\end{align}
holds from Eq.~(\ref{eq:z_MI_diagonalization}).
The unitarity of $2Z^{\mathrm{MI}} - I$ ensures $||Z^{\mathrm{MI}} - I/2||_2 = 1/2$, which leads to
\begin{align}
\max_{j,l} \left( \left| z^{\mathrm{MI}}_{j,l} - \frac{1}{2} \delta_{j,l} \right| \right) \leq
 \biggl\|Z^{\rm MI}-\frac{1}{2}I\biggr\|_2
 = \frac{1}{2}.
\end{align}
In particular,
considering the case $j \neq l$ and 
recalling
that all elements of $Z^{\mathrm{CDW}}$ are embedded in $Z^{\mathrm{MI}}$ as discussed in Sec.~\ref{sec:remarks_on_matrix_Z}, we obtain
\begin{align}
\max_{j,l,j\neq l} \left( \left| z^{\mathrm{MI}}_{j,l} \right| \right) \leq \frac{1}{2},\quad
\max_{j,l,j\neq l} \left( \left| z^{\mathrm{CDW}}_{j,l} \right| \right) \leq \frac{1}{2}.
\end{align}

The function $\max_{l=1}^{N} ( |z_{j,l}|, |\delta_{j,l} - z_{j,l}| )$ in the
definition of $\tilde{s}(\tilde{L})$ in Eq.~(\ref{eq:stilde_definition}) always
picks up the diagonal part of the matrix $Z$ because either $z_{j,j}$ or $1 - z_{j,j}$ is greater than $1/2$.
Therefore, 
we get
\begin{align}
\max_{l=1}^{N} ( |z_{j,l}|,|\delta_{j,l} - z_{j,l}| ) = \left| z_{j,j} - \frac{1}{2} \right| + \frac{1}{2}.
\label{eq:max_in_tilde_s}
\end{align}
Substituting Eq.~(\ref{eq:max_in_tilde_s}) into Eq.~(\ref{eq:stilde_definition}), we immediately obtain Eq.~(\ref{eq:stilde_simple_form}).

Next, we show
why the entropy-density-like values $\tilde{s}(\tilde{L})$ of the MI initial state
and 010101$\cdots$ CDW initial state with the same system size $L$ coincide.
To this end, we first examine the relation between the matrices $Z$ of the MI and CDW states.
By definition, even index elements of $Z^{\mathrm{MI}}$ are simply related to elements of $Z^{\mathrm{CDW}}$ through
\begin{align}
z^{\mathrm{MI}}_{2j,2l} = z^{\mathrm{CDW}}_{j,l} \quad (j,l = 1,2,\dots,L/2).
\label{eq:Z_MI_even_indices}
\end{align}
Note that the sizes of $Z^{\mathrm{MI}}$ and $Z^{\mathrm{CDW}}$ are, respectively, $L \times L$ and $L/2 \times L/2$.
Odd index elements of $Z^{\mathrm{MI}}$ are also 
related to elements of $Z^{\mathrm{CDW}}$.
The eigenfunction of the single-particle Hamiltonian $x_{k,j}$ is also the eigenfunction of the parity operator, satisfying $x_{k,L+1-j} = (-1)^{k+1} x_{k,j}$.
This leads to $y_{L+1-j,l} = y_{j,L+1-l}$ and
\begin{align}
z^{\mathrm{MI}}_{j,l} + z^{\mathrm{MI}}_{L+1-j,L+1-l} = \delta_{j,l},
\label{eq:z_MI_reflectivity}
\end{align}
where we assume that $L_{\mathrm{A}} = L/2$.
Using Eqs.~(\ref{eq:Z_MI_even_indices}) and (\ref{eq:z_MI_reflectivity}), 
we obtain
\begin{align}
z^{\rm MI}_{2j-1,2l-1} = \delta_{j,l} - z^{\rm CDW}_{L/2+1-j,L/2+1-l},
\label{eq:Z_MI_odd_indices}
\end{align}
for $j,l = 1,2,\dots,L/2$.
Using the relations between the matrices $Z$ of the MI and CDW states,
Eqs.~(\ref{eq:Z_MI_even_indices}) and (\ref{eq:Z_MI_odd_indices}), and
dividing the summation of $\tilde{s}(\tilde{L})$ in Eq.~(\ref{eq:stilde_simple_form}) of the MI initial state into the
even and odd index parts, we conclude
$\tilde{s}^{\mathrm{MI}}(\tilde{L})=\tilde{s}^{\mathrm{CDW}}(\tilde{L})$.

The coincidence of $\tilde{s}(\tilde{L})$ between the MI and 010101$\cdots$ CDW initial states seen in Figs.~\ref{fig:s2_div_L_cmp_hinf} and \ref{fig:s2_div_L_cmp_hinf_smaller} is rather special.
This is true only for the MI and 010101$\cdots$ CDW initial states with $L_{\mathrm{A}} = L / 2$.
When either condition is broken, such as considering 001001$\cdots$ CDW states or $L_{\mathrm{A}} = L / 4$, the coincidence disappears.

\bibliographystyle{apsrev4-2}

\onecolumngrid

\end{document}